\documentclass[pra,twocolumn,a4paper,showpacs,floatfix]{revtex4}
\usepackage{latexsym,amssymb,amsmath,graphicx}
\begin{document}

\title{Optical qudit-type entanglement
creation at long distances by means of small cross-Kerr
nonlinearities}
\date{\today}
\author{S. Ya. Kilin, A. B. Mikhalychev}
\affiliation{B. I. Stepanov Institute of Physics NASB, Minsk,
Belarus} 
\begin{abstract}
Entanglement represents an important resource for quantum
information processing, but its generation itself requires physical
resources that are limited. We propose a scheme for generating a
wide class of entangled qudit-type states of optical field modes at
sites separated by noisy medium when only weak optical
nonlinearities are available at both sites. The protocol is also
based on exploiting a weak probe field, transmitted between the
sites and used for generation of quantum correlations between two
spatially separated field modes. The idea of probabilistic
entanglement enhancement by measurement is discussed, and
corresponding scheme for measuring the probe field state with linear
optics and photodetectors not resolving photon numbers is proposed.
It is shown that the protocol is applicable in the case when
decoherence, limited efficiency and dark counts of photodetectors,
and uncertainty of nonlinear coupling constants are present.
\end{abstract}
\maketitle

\section{Introduction}

One of the most intriguing features of quantum mechanical
description of physical objects consists in presence of quantum
superpositions and especially of entangled states \cite{schr}. At
the very beginning of the quantum theory development such states
were found to possess quite counterintuitive (from the classical
point of view) properties \cite{EPR,schr_cat}. During the 20th
century the attitude of physicists to such quantum states
evolved from perceiving them as either evidences of theory
incompleteness or interesting but quite useless features of the
world \cite{schr,schr_cat,EPR,Bohr_EPR,Bohr_Einstein} to
understanding the opportunities provided by such quantum objects for
solving numerous tasks of information processing
\cite{BB84,E91,E91-modified,deutsch-1989,shor-1994,grover-1996}. It
has been shown that quantum no-cloning theorem
\cite{no-cloning-Zurek,no-cloning-dieks} provides unconditional
security of quantum cryptography protocols
\cite{BB84,E91,E91-modified,q_comp_communic,q_inf-cont_var,
lutkenhaus-2006}, while using entangled states of quantum register
can lead to essential speed-up of solving several classes of
complicated computational tasks
\cite{deutsch-1989,shor-1994,grover-1996,q_comp_communic,qi_processing,kilin1}.

From this point of view entanglement represents an important
resource for different information tasks. On the other hand,
entanglement itself requires some physical resources for its
generation and, thus, can be considered as an intermediate step on
the way from physical devices to accomplishing tasks of information
processing. Similarly to many other situations, the resources
available nowadays for entanglement generation are limited, and one
of the most important problems is to find methods for obtaining
results as good as possible using as few resources as possible.

Among systems, being promising for efficient entanglement
generation, optical field modes take their place due to possibility
of long-distance transmission with relatively low decoherence and
quite simple manipulation of the states with linear optics and
photodetectors. Considering optical entanglement generation, one can
divide necessary elements into two classes: quite simple elements,
available "freely" (linear optical devices, photodetectors not
resolving photon number, classical optical states), and
\emph{resources} --- all other elements, being quite challenging for
construction. It is impossible to generate entanglement with the
first group of elements only (see e.g.
Refs.~\cite{kim-2002-65,xiangbin-2002-66,asboth-2005-94}), and,
therefore, some resources are necessarily required. The main
resource, used for entanglement generation by optical methods, is
nonlinearity, either measurement-induced
\cite{ralph-2003,knill-2001} or provided by interaction between the
field modes in some medium. The corresponding physical resources are
special detectors (e.g. resolving photon number
\cite{ralph-2003,knill-2001}) and nonlinear media respectively. Entangled states can be created also by means of linear optics from
nonclassical states, which in their turn require nonlinearity for
their generation.

Many quantum information processing and communication tasks (quantum
cryptography, distributed quantum computation, teleportation of
quantum states) require distribution of entanglement between
parties, separated by noisy medium. For such tasks a high-quality
quantum channel also becomes an important physical resource. So, the
two main resources, required for generating entanglement between
distant states with optical methods, are nonlinear interaction and
a quantum channel. Therefore, efficient entanglement generation
corresponds in this case to creating strongly entangled states with
weak nonlinear interactions and noisy quantum channels (only such
systems are available nowadays).

In present paper we consider weak local cross-Kerr interaction as a
resource for creating nonlocal nonclassical states of spatially
separated optical field modes. Certain progress has already been
achieved in this field of research
\cite{vanloock-2006-96,louis-2007-9,louis-2007-75,spie,van_Loock-2008,
rohde-2008-8,nonlin_phen,opt_spect_eng}, especially in the case of
creating qubit-like entangled states. However, it remained quite
challenging to create more general classes of entangled states with
such limited resources. We propose a protocol for creating wide
class of qudit-type states (including entangled states) with
arbitrary dimensionality in continuous variable (optical) system
using weak cross-Kerr nonlinearity (as the main physical resources),
linear beamsplitters, detectors not resolving photon numbers, and
sources of coherent states. We show that entanglement of the states,
created with our protocol, can be higher than unity (which is the
limit for qubit-type states, that can be created by previously
proposed methods) and, thus, our protocol provides more effective
use of limited physical resources.

This paper is organized as follows. In the next section we discuss
the main ideas of entanglement generation between distant sites when
local nonlinear interaction and non-ideal quantum channel are
available. Then the main operations of the proposed protocol and
corresponding state transformations are presented.
Section~\ref{sec:discrim} is devoted to discussion of the
peculiarities of projecting the "raw" weakly entangled system state
onto strongly entangled desired final state. It is shown that
parameters of the protocol are determined in the unique way by the
desired final state, and corresponding relations are found. In
Section~\ref{sec:special} we demonstrate several applications of the
protocol to creation of nonclassical and entangled states of optical
field modes. In the last section we prove applicability of the
protocol for entanglement generation under realistic conditions by
taking into account decoherence, limited efficiency and dark counts
of photodetectors, and uncertainty of nonlinear coupling constants.

\section{Entanglement generation with local cross-Kerr nonlinearity}

\begin{figure}[t]
\begin{center}
\begin{tabular}{cc}
\textbf{(a)} & \\
& \includegraphics[scale=0.50]{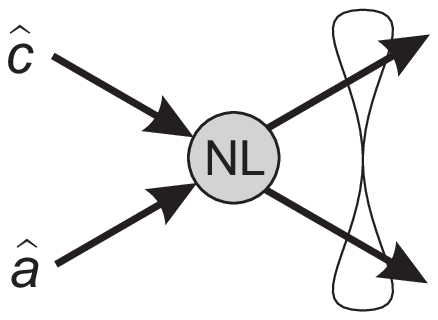}  \\ \\
\textbf{(b)} & \\
& \includegraphics[scale=0.50]{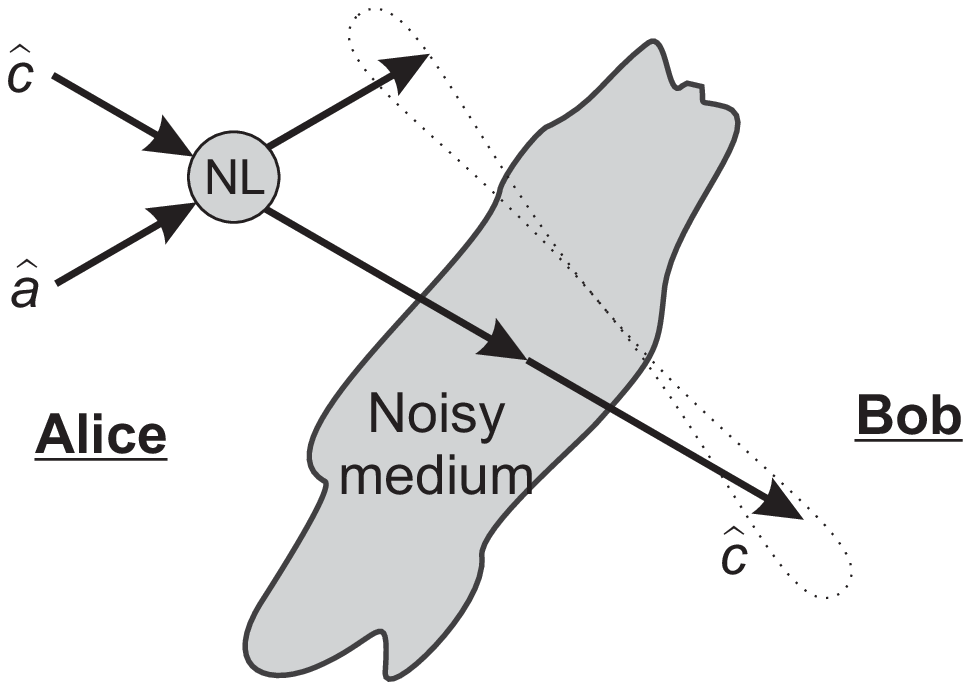}\\ \\
\textbf{(c)} & \\
& \includegraphics[scale=0.50]{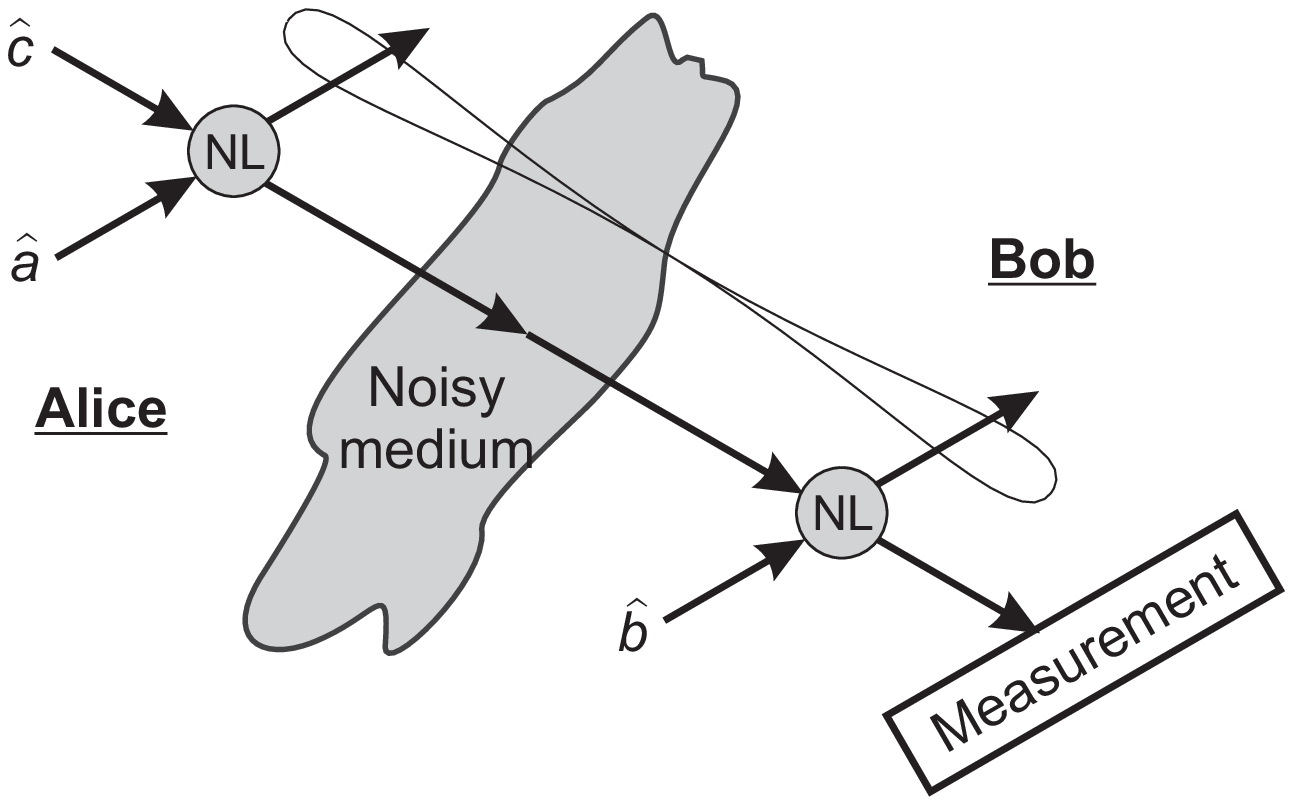}\\
\end{tabular}
\caption{\textbf{(a)} Entanglement generation by nonlinear
interaction. \textbf{(b)} Entanglement generation between sites,
separated by noisy medium. Only weak entanglement can be generated.
\textbf{(c)} Entanglement generation between sites, separated by
noisy medium, with probabilistic entanglement enhancement. Strongly
entangled states are available in the case of successful measurement
outcome.} \label{fig0}
\end{center}
\end{figure}

Cross-Kerr interaction itself can be used for generating entangled
states starting from uncorrelated states of a pair of quantum
objects. Suppose a field mode $\hat c$ in a coherent state $|\gamma
\rangle_c$ interacts with another system (another optical field mode
$\hat a$
\cite{spie,nonlin_phen,opt_spect_eng,tyc_korolkova,mogilevtsev_korolkova}
or an atomic system
\cite{vanloock-2006-96,van_Loock-2008,vanLoock-2006-8}). Due to the
interaction the phase of the coherent state amplitude $\gamma$ of
the mode $\hat c$ is shifted by the value, proportional to the
number of excitations $n$ of the system $\hat a$:
\begin{equation}
\label{eqn27} |n\rangle_a |\gamma \rangle_c \rightarrow |n\rangle_a
|\gamma e^{i \chi n}\rangle_c ,
\end{equation}
where $\chi$ describes effective strength of the interaction. If the
initial state of the object $\hat a$ is a superposition of states
with different excitation numbers (e.g. a coherent state $|\alpha
\rangle_a$ in the case of field mode), the final state of the
considered system will be an entangled state, composed by pairwise
combinations of coherent states with different phases of mode $\hat
c$ and number states of the system $\hat a$ (Fig.~\ref{fig0}(a)).

However, experimentally observed nonlinear interactions are quite
weak \cite{toulouse, kerr_fiber1,kang_zhu,kang_zhu-1}: the effective
nonlinearity strengthes, predicted in the most promising 4-level
atomic systems with electromagnetically induced transparency on the
basis of theoretical calculations and experimental data, have the
order of $\chi \sim 10^{-3}\div 10^{-2}$ \cite{imamoglu1, imamoglu2,
sinclair,kang_zhu,kang_zhu-1,li_yang-2008}. The magnitude of phase
space displacement caused by such cross-Kerr interaction is
proportional to $|\gamma| \chi$, and in general case the final state
can be weakly entangled. One of the solutions of the problem
consists in effective nonlinearity enhancement by using intense
fields $\hat c$: for $|\gamma| \gg 1$ the displacement magnitude can
be large enough ($|\gamma | \chi \gtrsim 1$) even for small $\chi$
\cite{vanloock-2006-96,louis-2007-9,louis-2007-75,vanLoock-2006-8}.

Entanglement generation gets more complicated if we take into
account not only limited available nonlinearity, but also noisy
medium between the sites (Fig.~\ref{fig0}(b)). Then decoherence
strongly limits maximal possible amplitudes of the transmitted field
$\hat c$ ($|\gamma| \ll 1$), while the amplitude of field $\hat a$
or the number of excitations in the atomic system is also limited by
losses in local storage. Therefore, this simple scheme is not
applicable for generating strongly entangled states when the sites
are separated by noisy media. Hence, under realistic conditions a more
sophisticated scheme, including some kind of entanglement
enhancement, is required.

One way of obtaining strongly entangled quantum state is to
implement entanglement distillation
\cite{ent_distill-76-722,ent_distill-54-3824,ent_distill-59-4206,
ent_distill-34-42-307}. This approach requires storage of quite a
large number of initial weakly entangled states and can be quite
challenging. Another solution of the problem can be based on
probabilistic entanglement enhancement \cite{spie,van_Loock-2008,
nonlin_phen,opt_spect_eng}. For this purpose one can design certain
measurement, carried out at Bob's site, with successful outcome
transforming initial weakly entangled state into a strongly
entangled state. Direct measurement of the state of field mode $\hat
c$ makes this mode inaccessible for any further use. Measurements,
implemented with linear optics and photodetectors not resolving
photon number on field modes, obtained by splitting the mode $\hat
c$, completely determine the state of the mode and destroy
entanglement. Therefore, some kind of nonlinearity is required at
Bob's site, too. It is quite natural to suppose that this nonlinear
interaction is the same as the one at Alice's site (see e.g.
Refs.~\cite{spie,van_Loock-2008, nonlin_phen,opt_spect_eng}).
%

Several schemes, based on probabilistic entanglement enhancement by
a measurement at Bob's site, have already been proposed for
entangling distantly separated field modes, when effective strength
of nonlinear interactions $\chi$ is equal to $\pi$ (strong nonlinear
interaction) \cite{spie}, and for entangling atomic qubits
\cite{van_Loock-2008} as well as for creating qubit-type entangled
states of optical field modes \cite{nonlin_phen,opt_spect_eng}, when
only small nonlinearity is available.


In present paper we solve a more general task of designing the
measurement scheme for creation of arbitrary qudit-type state
(including entangled states) from a wide class of possible states in
continuous variable system using weak cross-Kerr nonlinearity,
linear optical devices, detectors and sources of coherent states.
The set of achievable final states of the field modes $\hat a$ and
$\hat b$ has the form of a sum of phase-correlated pairs of coherent
states of the modes:
\begin{equation}
\label{eqn3} \left| \Psi_f \right \rangle_{ab} = \sum \limits _{n}
c_n \left|{\alpha e^{i\chi n}}\right\rangle_a \left|{\beta e^{i\chi
n}}\right\rangle_b,
\end{equation}
where coefficients $c_n$ are arbitrary and can be fixed in an
appropriate way for obtaining the state, most useful for certain
practical applications.

\section{Operations of the protocol}

\begin{figure*}[t]
\begin{center}
\includegraphics[scale=0.55]{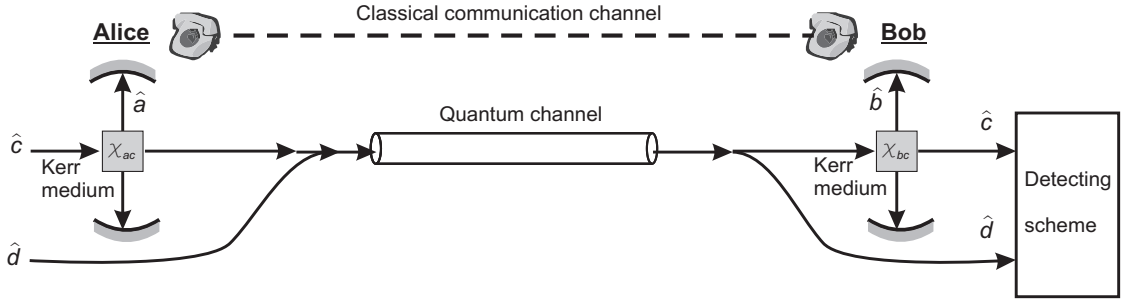}
\caption{Scheme of entanglement generation: Alice prepares entangled
state of the field modes $\hat a$ and $\hat c$; then she sends the
ancillary mode $\hat c$ and the reference mode $\hat d$ through the
quantum channel to Bob; having correlated the states of the modes
$\hat b$ and $\hat c$, Bob measures the final state of the ancillary
field mode (with the help of the reference mode $\hat d$) and, in
the case of successful outcome, announces Alice under the classical
channel that entanglement was generated (otherwise the set of
operations is repeated).} \label{fig1}
\end{center}
\end{figure*}

We consider the following system (see Fig.~\ref{fig1}): Alice and
Bob posses local field modes $\hat a$ and $\hat b$ referred to below
as main field modes; the probe beam (ancillary mode) is denoted as
mode $\hat c$; mode $\hat d$ is a reference field, transmitted from
Alice to Bob immediately before (or after) ancillary field for
decreasing influence of dephasing in the quantum channel on the
final states.

Main field modes, the ancillary and the reference fields are
prepared in coherent states ($|\alpha \rangle_a$ for Alice's mode
$\hat a$, $|\beta \rangle_b$ for Bob's mode $\hat b$, $|\gamma
\rangle_c$ for the probe field $\hat c$ and $|\tilde\gamma
\rangle_d$ for the reference field $\hat d$). The initial state of
the system is therefore uncorrelated one. Local cross-Kerr
interaction of the modes $\hat a$ and $\hat c$ (with effective
strength $\chi$ --- the phase of a coherent state of mode $\hat a$
increases by $\chi$ radians per each photon in $\hat c$) leads to
generation of correlations between the number of photons in the mode
$\hat c$ and the phase of the coherent state of the mode $\hat a$.
Then the field $\hat c$ is transmitted to Bob's site through the
quantum channel. In this section we suppose for simplicity that all
correlations are preserved by the channel. After local interaction
of the modes $\hat b$ and $\hat c$ (the effective strength of the
interaction is supposed in this section to be also equal to $\chi$),
taking place afterwards, the phases of the main modes $\hat a$ and
$\hat b$ become correlated with the number of photons in the mode
$\hat c$:
\begin{equation}
\label{eqn1}  \left|{\Psi_1}\right\rangle_{abc} =\sum\limits_{n} Q_n
(\gamma) \left|{\alpha e^{i\chi n}}\right\rangle_a \left|{\beta
e^{i\chi n}}\right\rangle_b \left|{n}\right\rangle_c,
\end{equation}
where $Q_n (\gamma) = \frac{\gamma^n}{\sqrt{n!}} e^{-|\gamma|^2/2}$.
However, the correlations, generated in the system, are weak in a
realistic case. For small ancillary field amplitudes $ |\gamma|^2
\ll 1$, required for decreasing decoherence in the quantum channel
\cite{spie,van_Loock-2008,opt_spect_eng}, the entanglement between
the states of the mode $\hat a$ possessed by Alice and the modes
$\hat b$ and $\hat c$ possessed by Bob is much less than unity:
\begin{equation}
\label{eqn2} E_1 \simeq h \left( \chi^2 |\alpha | ^2 |\gamma | ^2
\right) \ll 1,
\end{equation}
where $h(x) = -x \log_2 x - (1-x) \log_2 (1-x)$.

As discussed in the previous section, the last stage of the protocol
consists in detection of the probe beam state. This operation should
lead (in the case of successful outcome) to transformation of the
weakly correlated 3-modes state into a strongly correlated 2-modes
state of the form (\ref{eqn3}).

The idea of obtaining the desired strongly correlated final state
(\ref{eqn3}) from the state (\ref{eqn1}) is based on the following
decomposition:
\begin{equation}
\label{eqn4} \left|{\Psi_1}\right\rangle _{abc} =
\left|{\Psi_f}\right\rangle _{ab} \otimes
\left|{\varphi}\right\rangle _{c}  + \left|{\Psi_\bot}\right\rangle
_{abc},
\end{equation}
where $ \left|\Psi_\bot\right\rangle_{abc}$ denotes the part of the
system state, orthogonal to $|\varphi \rangle_c$, and the vector
$|\varphi \rangle_c$ is uniquely defined as
\begin{equation}
\label{eqn5}  \left|{\varphi}\right\rangle_c = const \cdot \sum
\limits_{n} \left( c_n^\ast / Q_n^\ast (\gamma) \right)
\left|{n}\right\rangle_c.
\end{equation}
Thus, the transformation of the state (\ref{eqn1}) into the state
(\ref{eqn3}) for arbitrary coefficients $c_n$ can be realized by
successful discrimination of the state $|\varphi \rangle_c$ from a
set of ancillary mode states, containing $|\varphi \rangle_c$ and a
complete system of states, orthogonal to $|\varphi \rangle_c$.

The most simple from theoretical point of view way of discriminating
the state $|\varphi \rangle_c$ is implementing projective
measurement described by operator $\hat P_\varphi =
\left|{\varphi}\mathrel{\left\rangle{\vphantom{\varphi}_c{}_c}\right\langle
\kern-\nulldelimiterspace}{\varphi}\right|$ which satisfies the
following relation:
\begin{equation}
\label{eqn6} \hat P_\varphi  \left|{\Psi_1}\right\rangle_{abc} =
\left|{ \Psi_f}\right\rangle_{ab} \otimes |\varphi \rangle_c.
\end{equation}

However, implementation of such measurement without additional
\emph{resources} (e.g. nonlinear interaction) can be too complicated
or impossible. Solution of this problem is discussed in the next
section, and the discrimination technique, based on a special case
of general class of POVM measurements --- "elimination"
measurements, --- is proposed.

\section{Discrimination technique}
\label{sec:discrim}

\subsection{"Elimination" measurements. General consideration}

An important fact that can be used for designing required
discrimination protocol is that coherent states represent a natural
basis for linear optical devices (the set of coherent states is
closed under linear optical transformations). From this point of
view, it is convenient to base the discrimination scheme on
comparing the state of the probe beam $\hat c$ with certain coherent
states, obtained by splitting the reference mode $\hat d$. However,
coherent states are nonorthogonal. It is this problem that leads to
impossibility of implementing the discussed above projective
measurement in a general case with linear optics. An important set
of measurements, implementable with linear beamsplitters and
photodetectors for coherent states, is the set of "elimination"
measurements (see e.g. Refs.~\cite{vanenk-2002-66,van_Loock-2008}
and Fig.~\ref{fig2}(a) below) --- the measurements with successful
outcomes manifesting that the field mode state \emph{is not} certain
coherent state.

The concept of measurements with outcomes, manifesting that the
input state \emph{is not} certain fixed state, was proposed in
Refs.~\cite{ivanovic,dieks,peres} as a part of unambiguous
discrimination of two non-orthogonal states. The concept of
"elimination" measurements was generalized in
Ref.~\cite{vanenk-2002-66} for the case of discrimination between
$N$ symmetric coherent states on a circle on phase plane with linear
optical elements and photodetectors. In these papers the system was
supposed to be prepared in \emph{one} of the states from a fixed
\emph{finite} set (e.g. symmetric coherent states on a circle:
$|\alpha e^{2 \pi i k / N} \rangle$, $k = 0, ..., N-1$).
Discrimination of one of the states (e.g. $|\alpha \rangle$) is
equivalent to \emph{elimination} of all the remaining states from
the fixed set (in the considered example $|\alpha e^{2 \pi i k / N}
\rangle$, $k = 1, ..., N-1$).

Considering ancillary mode states in our protocol, one faces a more
general situation. A complete set of independent states is infinite
for a field mode, and the measured state does not necessarily
coincide with certain basis vector, but can represent a
superposition of basis vectors. Therefore, complete description of
state transformations, occurring when elimination measurements are
carried out, requires a more rigorous operator definition of the
concept of elimination measurements.


We define \emph{measurement, eliminating state $|\psi \rangle$,} as
any POVM-type measurement (described by POVM $\{\hat A_i  \}$, $\sum
_i [\hat A_i ]^+ \hat A_i  =1$) with at least one of the outcomes
(referred to in this paper as successful) being characterized by
operator, denoted in further consideration as $A_{|\psi \rangle }$,
with the property
\begin{equation}
\label{eqn7} \hat A_{|\psi \rangle }|\psi \rangle = 0.
\end{equation}
Having obtained this outcome, one can with certainty conclude that
the measured state \emph{was not} the state $|\psi \rangle$.

Considering a set of measurements eliminating states $\left\{|\psi_j
\rangle \right\}$, we will require (simultaneously with
Eq.~(\ref{eqn7}) for all vectors $|\psi_j \rangle$) commutativity of
operators, characterizing successful outcomes of the measurements:
\begin{equation}
\label{eqn23} \left[ \hat A_{|\psi_i \rangle }, \hat A_{|\psi_j
\rangle }  \right]=0.
\end{equation}
This condition means that successful elimination of the state
$|\psi_i \rangle$ must not destroy the result of elimination of the
state $|\psi_j \rangle$ for any pair of the states $|\psi_i \rangle$
and $|\psi_j \rangle$ from the considered set.

The conditions (\ref{eqn7}), (\ref{eqn23}) lead to the following
important implication, useful for designing the required
discrimination scheme. Let the set $\left\{ |\psi_0 \rangle, ...,
|\psi_K \rangle \right\}$ be a (non-orthogonal) basis of
finite-dimensional Hilbert space and operators $\hat A_{|\psi_0
\rangle}$, ..., $\hat A_{|\psi_K \rangle}$ describe successful
outcomes of corresponding elimination measurements. Any state $|\phi
\rangle$ from the considered state space
\begin{equation}
\label{eqn28} |\phi\rangle = \phi_0 |\psi_0 \rangle + ... + \phi_K
|\psi_K\rangle,
\end{equation}
subjected to successful elimination of the states $\left\{ |\psi_1
\rangle, ..., |\psi_K \rangle \right\}$, is transformed into the
state
\begin{equation}
\label{eqn30}\begin{aligned} \hat A_{|\psi_K \rangle} &... \hat
A_{|\psi_1 \rangle}|\phi\rangle ={}\\{}={}& \phi_0 \hat A_{|\psi_K
\rangle} ... \hat A_{|\psi_1 \rangle} |\psi_0 \rangle,
\end{aligned}
\end{equation}
where contribution of $|\psi_0 \rangle$ only did not vanish. The
final state of the system differs from the state $|\psi_0 \rangle$,
but it is a completely defined state and can be transformed
unitarily into $|\psi_0 \rangle$ (in our protocol the ancillary mode
is finally discarded and this transformation is not necessary).
Therefore, one can state that successful elimination of all the
basis vectors except $|\psi_0 \rangle$ leads to successful
\emph{discrimination} of the state $|\psi_0 \rangle$.


Generalization of the definition to the case, when several operators
$\hat A _{|\psi_j\rangle }^{(\mu)}$, $\mu= 1, 2, ...$, correspond to
successful elimination of the state $(|\psi_j\rangle )$: $\hat A
_{|\psi_j\rangle }^{(\mu)}|\psi_j \rangle = 0$, is straightforward:
Eq.~(\ref{eqn23}) retains its form and must be valid for all $\hat A
_{|\psi_j\rangle }^{(\mu)}$, used instead of single operator $\hat A
_{|\psi_j\rangle }$. The result, described by Eq.~(\ref{eqn30}),
also remains valid.

\subsection{State discrimination with
"elimination" measurements. From infinite-dimensional to
finite-dimensional space}

The described above technique can be used exactly for discrimination
of the state $|\varphi \rangle_c$ of the ancillary mode $\hat c$ in
a special case of finite-dimensional space of input states. Suppose
that the final state of the whole system (modes $\hat a$, $\hat b$
and $\hat c$) can be decomposed using finite number of ancillary
field mode states, and these states together with the state
$|\varphi \rangle_c$ span $(K+1)$-dimensional space. Then we can
choose $K$ such independent vectors $|\psi_j \rangle_c$, $j=1,...,K
$, orthogonal to $|\varphi \rangle_c$, that the system $\left\{
|\psi_1 \rangle_c,..., |\psi_K \rangle_c, |\varphi \rangle_c
\right\}$ is complete in corresponding subspace. As shown above,
discrimination of the state $|\varphi \rangle_c$ corresponds to
successful elimination of the vectors $|\psi_j \rangle_c$,
$j=1,...,K $.

Such finite-dimensional case can be realized, for instance, when the
nonlinearity strength is equal to $\chi=2 \pi / N$, where $N$ is an
integer \cite{spie}. In these special case the state of the modes
$\hat a$, $\hat b$ and $\hat c$ after the nonlinear interactions can
be represented as a sum of $N$ terms consisting of coherent state of
the form $|\gamma e^{2 \pi k / N} \rangle_c$ of the ancillary mode
$\hat c$ and corresponding entangled state of modes $\hat a$ and
$\hat b$. Discrimination of a fixed state $|\gamma e^{2 \pi k_0 / N}
\rangle_c$ (by elimination of $N-1$ coherent states $|\gamma e^{2
\pi k / N} \rangle_c$, $k\ne k_0$) maps initial weakly-entangled
state onto strongly-entangled final state of the modes $\hat a$ and
$\hat b$.

In general infinite-dimensional case, however, the state of the
system cannot be decomposed in such a way and the described
discrimination technique can be applied approximately. For the
considered state (\ref{eqn1}) and for small ancillary field
amplitudes $|\gamma| \ll 1$ high accuracy of exploiting the
discrimination technique can be achieved by restricting
consideration by a space of states with limited photon numbers.
Suppose the expression (\ref{eqn3}) for the desired final state has
$K+1$ nonzero terms, i.e. $c_n = 0$ for $n>K$. Then Eq.~(\ref{eqn5})
for the state $|\varphi \rangle_c$ also has $K+1$ nonzero terms and
this state belongs to $(K+1)$-dimensional subspace of states with
limited photons numbers $n \le K$. According to Eq.~(\ref{eqn1}),
the probability of presence of more than $K$ photons in the mode
$\hat c$ is proportional to $|\gamma|^{2(K+1)}$ and is small for the
considered system. Therefore, the state (\ref{eqn1}) with
probability close to unity belongs to the same $(K+1)$-dimensional
subspace as the state $|\varphi \rangle$:
\begin{equation}
\label{eqn36} \begin{aligned} |\Psi_1 \rangle_{abc} =
\sum\limits_{n=0}^K Q_n (\gamma) \left|{\alpha e^{i\chi
n}}\right\rangle_a \left|{\beta e^{i\chi n}}\right\rangle_b
\left|{n}\right\rangle_c +\\+ |\delta \Psi_1^{(K)} \rangle_{abc}   =
|\Psi_1^{(K)} \rangle_{abc} + |\delta \Psi_1^{(K)} \rangle_{abc},
\end{aligned}
\end{equation}
where $\|  |{\delta \Psi_1^{(K)}}\rangle_{abc}  \| =
O\left(|\gamma|^{K+1}\right)$. If we construct a measurement, which
leads to correct discrimination of the state $|\varphi \rangle_c$
for the subspace, spanned by the state vectors with photon numbers
$n \le K$, the distance between the obtained state (for the ideal
system) and the desired final state $\left| \Psi_f \right \rangle$
will have the order not greater than
\begin{equation}
\label{eqn37} \frac{\|  |{\delta \Psi_1^{(K)}}\rangle_{abc}  \|^2}
{\left\| \hat A_{|\psi_K \rangle_c} ... \hat A_{|\psi_1 \rangle_c}
|\Psi_1^{(K)} \rangle_{abc} \right\|^2 } =
\frac{O\left(|\gamma|^{2(K+1)}\right)} {p_K},
\end{equation}
where $p_K$ is the probability of successful generation of the
desired final state. Therefore, if the probability $p_K$ has the
order at least $O\left(|\gamma|^{2K}\right)$, the error of
exploiting the approximate discrimination technique will have the
order $ O\left( |\gamma|^{2} \right)$ which is much smaller than the
decrease of the final state fidelity caused by non-ideality of the
quantum channel. Calculations below show that this condition is
fulfilled for the considered system.

Thus, for generating the state $|\Psi_f \rangle_{ab}$ composed by
the sum of $K+1$ phase correlated pairs of coherent states of modes
$\hat a$ and $\hat b$, one needs to construct a scheme for
elimination of $K$ vectors $|\psi_j \rangle_c$, $j=1,...,K $.

\subsection{Coherent states as the basis set for discrimination}

The next step of designing the scheme for discrimination of the
state $|\varphi \rangle_c$ is construction of the set of independent
states $\left\{ |\psi _j \rangle_c \right\}$ in the form suitable
for realizing elimination measurements with linear optical elements
and photodetectors. Realization of such measurement for coherent
states is known (see e.g. Ref.~\cite{vanenk-2002-66}), therefore, it
is desirable that the basis states $ |\psi _j \rangle_c $ be
coherent states $|\gamma_j \rangle_c$ with amplitudes $ \gamma_j $.
In this case the orthogonality conditions have the form $
{\strut}_c\left\langle{\varphi}\mathrel{\left|{\vphantom{\varphi
\gamma_j}}\right.
\kern-\nulldelimiterspace}{\gamma_j}\right\rangle_c =\sum c_n
\gamma_j^n / \left( Q_n (\gamma) \sqrt{n!}\right) = 0$. Then the
coherent states amplitudes $\gamma_j $ can be found as $K$ roots of
the $K$-th order equation
\begin{equation}
\label{eqn8}   f(x)\equiv \sum\limits _{n=0}^K  c_n \left(
\frac{x}{\gamma} \right)^n = 0
\end{equation}
and are uniquely defined for any given set $\left\{ c_n \right\}$.
These statements are correct in the nondegenerate case.

Presence of degenerate roots of Eq.~(\ref{eqn8}) leads to linear
dependence of the set $\left\{|\gamma_j \rangle \right \}$. In this
case additional vectors must be added to the set to provide
completeness. These vectors can be constructed in the following way.
If the root $\gamma_m$ has the multiplicity $l_m
> 1$, then for $x=\gamma_m$ holds
\begin{equation}
\label{eqn9} \frac{d^s f(x)}{ d^s x} = 0,\qquad 0 \le s \le l_m -1.
\end{equation}
On the other hand,
\begin{equation}
\label{eqn10} \begin{aligned} \frac{d^s f(\gamma_m) }{ d^s \gamma_m}
&= \frac{d^s }{ d^s \gamma_m} \left\{
{\strut}_c\left\langle{\varphi}\mathrel{\left|{\vphantom{\varphi
\gamma_m}}\right.
\kern-\nulldelimiterspace}{\gamma_m}\right\rangle_c
e^{|\gamma_m|^2/2}\right\} =\\&=
{\strut}_c\left\langle{\varphi}\mathrel{\left|{\vphantom{\varphi
\gamma_m}} \left(\hat c^+ \right)^s \right|
\kern-\nulldelimiterspace}{ \gamma_m}\right\rangle_c
e^{|\gamma_m|^2/2}.
\end{aligned}
\end{equation}
Eqs.~(\ref{eqn9})--(\ref{eqn10}) prove orthogonality of the vectors
$|\gamma_m^{(s)} \rangle_c =  \left( \hat c^+ \right)^s | \gamma_m
\rangle_c $ to $|\varphi \rangle_c $ for $s\le l_m -1$. Instead of
$l_m$ copies of the vector $|\gamma_m \rangle_c$ we obtain $l_m$
vectors $ | \gamma_m^{(s)} \rangle_c$, $s=0,..., l_m -1$, that are
(i) independent and (ii) orthogonal to $|\varphi \rangle_c$.

Thus, in general case the set of vectors $|\gamma_m^{(s)} \rangle_c
= \left( \hat c^+ \right)^s | \gamma_m \rangle _c$, where $s=0, ...,
l_m-1$ and index $m$ enumerates distinct roots of Eq.~(\ref{eqn8}),
is the required set of $K$ independent vectors, orthogonal to
$|\varphi \rangle_c$. This set of states can be constructed for any
desired final state of the form (\ref{eqn3}) with finite number of
terms and is uniquely defined for a fixed set of coefficients
$\left\{ c_n \right\}$. Therefore, Eq.~(\ref{eqn8}) provides the
unique solution of the problem of generating any final state of the
form~(\ref{eqn3}) with minimal exploited resources.

\subsection{Implementation:
nondegenerate case}

\begin{figure}[t]
\begin{center}
\begin{tabular}{cc}
\textbf{(a)} & \\
& \includegraphics[scale=0.55]{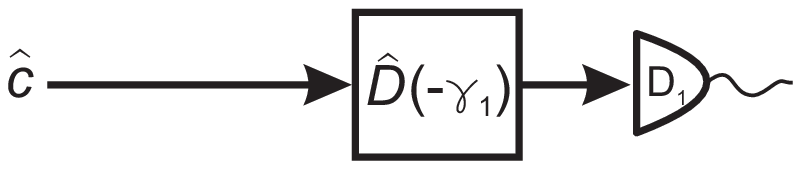}  \\ \\
\textbf{(b)} & \\
& \includegraphics[scale=0.55]{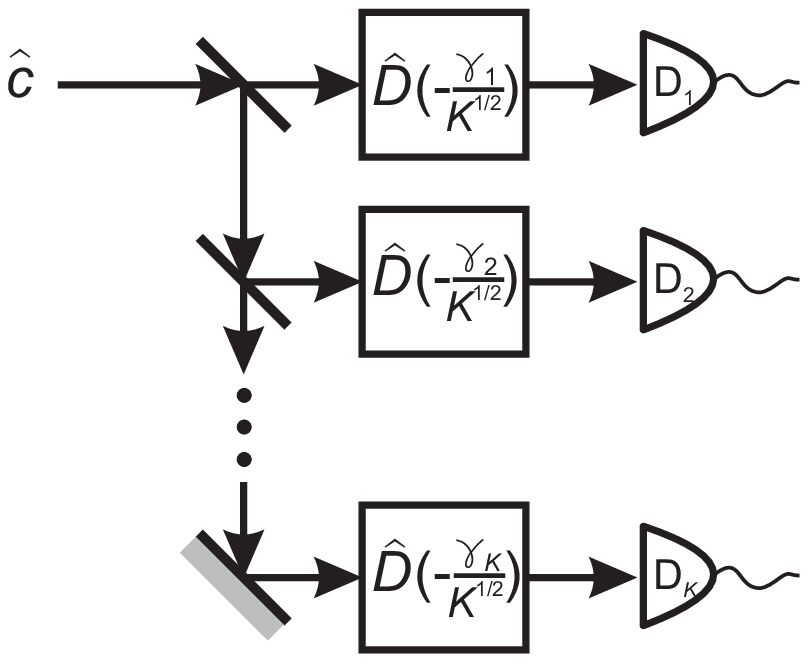}\\ \\
\textbf{(c)} & \\
& \includegraphics[scale=0.55]{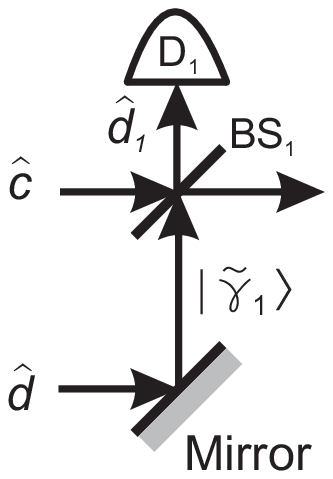}\\ \\
\textbf{(d)} & \\
& \includegraphics[scale=0.55]{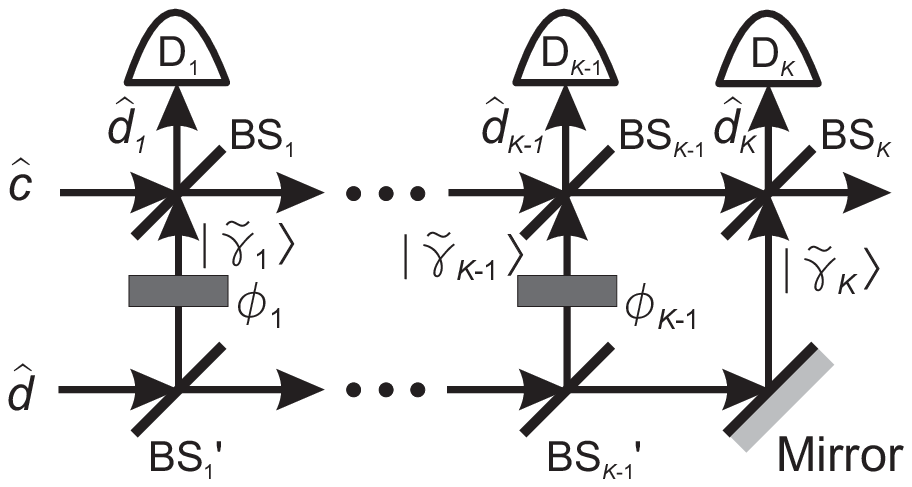}\\
\end{tabular}
\caption{\textbf{(a)} Elimination measurement for a single coherent
state $|\gamma_1 \rangle_c$. $\hat D_c (-\gamma _1)$ is the operator
of coherent displacement with the magnitude $-\gamma_1$.
\textbf{(b)} Scheme for elimination of a set of coherent states $
|\gamma_j \rangle_c$, $j= 1,...K$. \textbf{(c)} Elimination
measurement for a single coherent state $|\gamma_1 \rangle_c$ with
the coherent displacement operator $\hat D_c (-\gamma _1)$ being
effectively realized by mixing the mode $\hat c$ with the reference
mode $\hat d_1$ in coherent state $|\tilde \gamma_1 \rangle_{d_1}$
at linear beamsplitter $BS_1$. \textbf{(d)} Scheme for elimination
of coherent states $ |\gamma_j \rangle_c$, $j= 1,...K$, with linear
beamsplitters ($BS_1$, ..., $BS_{K-1}$; $BS_1'$, ..., $BS_{K-1}'$)
and photodetectors($D_1$,..., $D_K$); reference coherent states for
implementing necessary displacements in phase space are obtained by
splitting reference field $\hat d$ and applying phase shifts
$\phi_1$,~..., $\phi_{K-1}$.} \label{fig2}
\end{center}
\end{figure}

In the nondegenerate case discrimination of the state $|\varphi
\rangle_c$ is based on quite a well known technique of eliminating
coherent states (Fig.~\ref{fig2}).
For example, for eliminating a single coherent state $|\gamma_1
\rangle_c$ the measured state is displaced in phase space by the
magnitude $-\gamma_1$ (the displacement is described by operator
$\hat D_c (-\gamma_1 ) = \exp \left( -\gamma_1 \hat c^+
+\gamma_1^\ast \hat c\right)$). The displacement operator transforms
coherent state $|\gamma_1 \rangle_c$ into the vacuum state, and
detection of photons in the field mode after the displacement
(Fig.~\ref{fig2}(a)) manifests that the measured state \emph{was
not} $|\gamma_1 \rangle_c$. It should be noted that the considered
detectors need not resolve photon numbers or be 100\% efficient.

Elimination of a set of coherent states $\left\{|\gamma _j \rangle_c
\right\}$ can be carried out in a similar way by splitting the field
mode $\hat c$ into $K$ modes (Fig.~\ref{fig2}(b)). Each of the
obtained modes is used for eliminating one of the coherent states
$|\gamma_j \rangle_c$. In this case obtaining photocounts ("clicks")
from all the detectors corresponds to successful outcomes of
elimination of all the vectors $\left\{ |\gamma_j \rangle_c ,
j=1,...,K \right\}$ and, thus, to successful discrimination of
$|\varphi\rangle_c$.

The displacement operators can be effectively realized by mixing the
field mode $\hat c$ with additional reference modes $\hat d_1$,~...,
$\hat d_K$ in corresponding coherent states $|\tilde
\gamma_1\rangle_{d_1}$,~..., $|\tilde \gamma_K\rangle_{d_K}$
(prepared by splitting the reference mode $\hat d$ and applying
additional phase shifts) at linear beamsplitters
(Fig.~\ref{fig2}(c)). For example, for elimination of a single
coherent state $|\gamma_1 \rangle_c$ one mixes the mode $\hat c$
with a single reference mode $\hat d_1$ at linear beamsplitter
$BS_1$ with transmittance $T_1 = \cos^2 \theta_1$
(Fig.~\ref{fig2}(c)). If the amplitude of the reference mode
coherent state equals $\tilde \gamma_1  = - i \gamma_1 \tan
\theta_1$ and the measured state of the mode $\hat c$ is $|\gamma_x
\rangle_c$, after mixing at the beamsplitter the modes will be in
the state $|\cos \theta_1 \cdot (\gamma_x + \gamma_1 \tan^2
\theta_1) \rangle_c |i \sin \theta_1 \cdot (\gamma_x - \gamma_1)
\rangle_{d_1}$. The obtained state of the mode $\hat c$ can be used
for some further operations, while the mode $\hat d_1$ appears just
in the state, required for implementing elimination measurement: the
amplitude of the initial coherent state $|\gamma_x \rangle_c$ is
displaced by magnitude $-\gamma_1$. Additional phase factor $i$ is
irrelevant for detection of presence of photons in the mode, and
decrease of the field intensity by the factor $\sin^2 \theta_1 = 1 -
T_1$ influences only the probability of detecting photons and, thus,
of obtaining successful elimination outcome. In an important case,
when the mode $\hat c$ is discarded after the elimination
measurement, the efficiency of the measurement can be improved by
requiring $T_1 = \delta \ll 1$ (then $\sin^2 \theta_1 \approx 1$ and
success probability is approximately the same as for the scheme in
Fig.~\ref{fig2}(a)).

The scheme suitable for implementing elimination measurements in
general case of $K$ coherent states $\{|\gamma_j \rangle_c,
j=1,...,K \}$ is shown in Fig.~\ref{fig2}(d). If the measured state
of the mode $\hat c$ is coherent state $|\gamma_x \rangle_c$, we
require the states of the reference modes $\hat d_j$ after mixing
with the mode $\hat c$ at beamsplitters to be coherent states $|i q
( \gamma_x - \gamma_j) \rangle_{d_j}$, where $q$ is certain
coefficient independent of the mode number $j$. These states
correspond to displacement of the amplitude $\gamma_x$ of the
measured coherent state by magnitudes $\gamma_j$ and can be used for
carrying out corresponding elimination measurements.

Transmittances $T_j$ of the beamsplitters $BS_j$ in this case are
defined in the unique way by the above requirement of obtaining
correct coherent displacements. As in the case of single coherent
state $|\gamma_1 \rangle_c$, we require the transmittance of the
last beam splitter to be small: $T_K  = \delta \ll  1$ (the mode
$\hat c$ is discarded after implementing elimination measurements).
The condition of dividing the amplitude $\gamma_x$ of measured
coherent state into equal parts $q$ between the modes $\hat d_j$
leads to the following system of equations for the beamsplitter
transmittances (we define parameters $\theta_j$ by $T_j= \cos^2
\theta_j$):
\begin{equation}
\label{eqn67} \left\{
\begin{array}{l}
\sin \theta_1 = q, \\
\sin \theta _2 \cos \theta_1 = q, \\
\quad ...\\
\sin \theta_K \cos \theta _{K-1}\cdot ... \cdot \cos \theta_1 = q.
\end{array}
\right.
\end{equation}
Solving these equations together with the requirement $T_K = \delta
$, one finds the following expressions for the transmittances of the
beamsplitters:
\begin{equation}
\label{eqn35} T_j = \frac{(K-j-1)(1-\delta)+1 }{ (K-j)(1-\delta)+1}
\approx \frac{K-j} {K+1-j},
\end{equation}
and coefficient $q$:
\begin{equation}
\label{eqn72} q=\left(K+\frac{\delta}{1-\delta} \right)^{-1/2}
\approx 1/\sqrt{K}.
\end{equation}

The amplitudes of the reference modes coherent states $ \tilde
\gamma_m $ are also defined in the unique way by the requirement of
obtaining correct coherent displacements in implemented elimination
measurements. The following recurrent system of equation can be
obtained:
\begin{equation}
\label{eqn73} \left\{
\begin{array}{l}
\tilde \gamma_j \cos \theta_j +i  \check \gamma_{j-1} \sin \theta_j
=
-i q \gamma_j, \\
\check \gamma_{j} = \check \gamma_{j-1} \cos \theta _j + i
\tilde \gamma_j \sin \theta _j, \\
\check \gamma_0 =0,
\end{array}
\right.
\end{equation}
where $\check \gamma_j $ is the amplitude of reference coherent
state, mixed to the mode $\hat c$ by first $j$ beamsplitters. The
solution of this system of equations is
\begin{equation}
\label{eqn75} \tilde \gamma_j = - \frac{i q } {\cos \theta_j}
\left\{\gamma_j  +  \sin^2 \theta_j ( \gamma_1 + ... + \gamma_{j-1})
\right\}.
\end{equation}

As stated above, coherent states of reference mode $\hat d_j$ with
the required amplitudes $\tilde \gamma_j$ can be obtained by
splitting coherent state $\tilde \gamma$ of the mode $\hat d$
(Fig.~\ref{fig2}(d)). The transmittances $T_j' = \cos \theta_j '$ of
the beamsplitters $BS_j'$ and the phase shifts $\phi_j$ are
solutions of the following system of equations:
\begin{equation}
\label{eqn76}\begin{aligned} & \tilde \gamma_j = i \cos \theta_1'
\cdot ... \cdot \cos \theta_{j-1}' \sin \theta_j' e^{i \phi_j}
\tilde \gamma, \\ & \qquad \qquad \qquad \qquad j=1,...,K-1; \\
& \tilde \gamma_K = i \cos \theta_1' \cdot ... \cdot \cos
\theta_{K-1}' \tilde \gamma.
\end{aligned}
\end{equation}
The system contains $K$ complex equations for $2K$ real variables
$T_1'$,~..., $T_{K-1}'$, $\phi_1$,~..., $\phi_{K-1}$,
$\operatorname{Re} \tilde \gamma$, $\operatorname{Im} \tilde
\gamma$. The solution of the equations is quite cumbersome, and we
will provide it in explicit form only for certain special cases,
discussed below.

\subsection{Implementation:
degenerate case}

In the degenerate case we need to eliminate not only coherent
states, but also the states of the form $ |\gamma_m^{(s)}\rangle_c =
\left( \hat c^+ \right)^s | \gamma_m \rangle_c$, created from
coherent states by adding fixed number of photons (photons-added
coherent states --- PACS). We will show that elimination of PACS can
be done exactly in the same way as elimination of coherent states
$|\gamma_j \rangle_c$.

For designing a scheme, suitable for elimination of the coherent
state with one added photon, e.g. $\hat c^+ |\gamma_1 \rangle_c$, it
is useful to notice that when a beam with a single-photon excitation
is split into two parts by a beamsplitter, the excitation can be
detected in one of the two beams, but not in both of the beams
simultaneously.
If the mode $\hat c$, the state of which is measured, is initially
in the state $\hat c^+ |\gamma_1 \rangle_c$, after splitting the
mode into two parts (e.g. modes $\hat c_1$ and $\hat c_2$) we obtain
a superposition state $\frac{1}{\sqrt{2}} (\hat c_1^+ + \hat c_2^+ )
|\gamma_1 / \sqrt 2 \rangle_{c_1} |\gamma_1 / \sqrt 2
\rangle_{c_2}$, where the added photon can be found in one of the
modes, but never in the two modes simultaneously. Therefore, for all
the terms of the superposition at least one of the modes is in
coherent state $|\gamma_1 /\sqrt 2 \rangle$. If we implement
measurements, eliminating the state $|\gamma_1 /\sqrt 2 \rangle$,
for both of the modes, two successful outcomes can never be
obtained, if the measured state of the mode $\hat c$ was $\hat c^+
|\gamma_1 \rangle_c$. Therefore, two simultaneously obtained
successful outcomes of the elimination measurements correspond to
\emph{elimination} of the PACS with one added photon $\hat c^+
|\gamma_1 \rangle_c$. It should be noted that exactly the same
scheme would be obtained, if we tried to eliminate two coherent
states $|\gamma_1 \rangle_c$ and $|\gamma_2 \rangle_c$ with equal
amplitudes $\gamma_1 = \gamma_2$ by the method, suitable for
nondegenerate case.

Elimination of a PACS with $s$ added photons (e.g. $|\gamma_1^{(s)}
\rangle_c$) can be carried out in a similar way, taking into account
that when a mode with $s$ photons is split into $s+1$ parts, it is
impossible to detect a photon in each of the $s+1$ modes
simultaneously. Therefore, for elimination of the state
$|\gamma_1^{(s)} \rangle_c$ one can splits the mode $\hat c$ into
$s+1$ modes. When $s$ photons, added to coherent state $|\gamma_1
\rangle_c$ according to the definition of the state $|\gamma_1^{(s)}
\rangle_c$, are distributed between $s+1$ modes, at least one mode
appear in coherent state $|\gamma_1 /\sqrt{s+1} \rangle$ without
added photons. Therefore, $s+1$ successful outcomes of elimination
of coherent state $|\gamma_1 /\sqrt{s+1} \rangle$ for the $s+1$
modes cannot be obtained simultaneously if the initial state of the
mode $\hat c$ is the state $|\gamma_1^{(s)} \rangle_c$ (or any of
the states $|\gamma_1 \rangle_c$, $|\gamma_1^{(1)} \rangle _c$,~...,
$|\gamma_1^{(s-1)} \rangle_c$ with lesser numbers of added photons).
Thus, obtaining successful outcomes of the $s+1$ measurements,
eliminating coherent state $|\gamma_1 /\sqrt{s+1} \rangle$,
corresponds to elimination of the state $|\gamma_1^{(s)} \rangle_c$,
as well as of the states $|\gamma_1 \rangle_c$, $|\gamma_1^{(1)}
\rangle _c$,~..., $|\gamma_1^{(s-1)} \rangle_c$. As in the case,
discussed in the previous paragraph, exactly the same scheme could
be used for elimination of the set of coherent states $|\gamma_1
\rangle_c$,~..., $|\gamma_s \rangle_c$ with equal amplitudes
$\gamma_1= \gamma_2 =  ... = \gamma_s$ if the case were considered
as nondegenerate.

The full algorithm of designing discrimination scheme for degenerate
case can be summarized as follows. At first, $K$ roots $\gamma_j$ of
Eq.~(\ref{eqn8}) are found. Then the set of coherent states
$|\gamma_j \rangle_c$ is constructed. If the root $\gamma_m$ has
multiplicity $l_m>1$, $l_m$ "copies" of the coherent state
$|\gamma_m \rangle_c$ are replaced by $l_m$ independent states
$|\gamma_m \rangle_c$, $|\gamma_m^{(1)} \rangle_c$,~...,
$|\gamma_m^{(l_m-1)} \rangle_c$. In the obtained set of independent
states single coherent states are eliminated by the method,
discussed in the previous subsection. Sets of the states $|\gamma_m
\rangle_c$, $|\gamma_m^{(1)} \rangle_c$,~..., $|\gamma_m^{(l_m-1)}
\rangle_c$ with different numbers of photons, added to the same
coherent state, are eliminated by carrying out measurements,
eliminating coherent state $|\gamma_m \rangle$ (with amplitude,
decreased by splitting), for $l_m$ modes, obtained after splitting
the mode $\hat c$. Therefore, $l_m$ "copies" of the root $\gamma_m$,
appearing in the list of roots of Eq.~(\ref{eqn8}), correspond in
the final discrimination scheme to $l_m$-fold elimination of the
coherent state $|\gamma_m \rangle_c$. It means that one need not
make any difference between degenerate and nondegenerate roots of
Eq.~(\ref{eqn8}), eliminating coherent states $|\gamma_j \rangle_c$
as many times, as they appear in the list of roots of
Eq.~(\ref{eqn8}). Thus, the scheme, designed for discrimination of
the state $|\varphi \rangle_c$ in nondegenerate case
(Fig.~\ref{fig2}(d)), is also suitable for degenerate case.

\subsection{Implementation:
mathematical description}

Quite interesting result of applicability of the same scheme for
both nondegenerate and degenerate case can be given more rigorous
mathematical proof on the basis of operator definition of
elimination measurements (Eqs.~(\ref{eqn7}), (\ref{eqn23})).

As shown in Appendix~\ref{app1} (see Eq.~(\ref{eqn49a})),
transformation of the system density matrix in the case of
successful outcomes of the $K$ measurements, eliminating coherent
states $|\gamma_j \rangle_c$, (photocounts obtained from all the
detectors $D_j$) has the form
\begin{equation}
\label{eqn49}\begin{aligned} \rho_{abc}^{(out)} = \hat M \Bigl\{
\sum_{n_j\ge 1} \hat A^{(n_K)} _{|\gamma_K \rangle_c} ...  \hat
A^{(n_1)}_{ |\gamma_1 \rangle_c} \rho_{abc}^{(in)} \otimes \\
\otimes \left( A^{(n_1)}_{ |\gamma_1 \rangle_c}\right)^+ ...
\left(A^{(n_K)}_{ |\gamma_K \rangle_c}\right)^+ \Bigr)\}
\end{aligned}
\end{equation}
where density matrices $\rho_{abc}^{(in)}$ and $\rho_{abc}^{(out)}$
describe the system state before and after implementing elimination
measurements respectively; superoperator $\hat M$
(Eq.~(\ref{eqn32})) describes the part of system state
transformation, which does not depend on measurement outcomes; the
sets of operators $\{ \hat A^{(n_j)} _{|\gamma_j \rangle_c}, n_j =
1,2,... \}$ correspond to successful elimination of coherent states
$|\gamma_j \rangle_c$:
\begin{equation}
\label{eqn50} \hat A^{(n_j)} _{|\gamma_j \rangle_c} =
\frac{q^{n_j}}{\sqrt{n_j!}} \left( \hat c - \gamma_j \right)^{n_j},
\end{equation}
\begin{equation}
\label{eqn51} \hat A^{(n_j)} _{|\gamma_j \rangle_c}
|\gamma_j\rangle_c=0, \quad n_j=1,2,...
\end{equation}
and satisfy Eq.~(\ref{eqn23}).

The scheme is apparently suitable for nondegenerate case, and one
needs to shows that it also can be used when some roots of
Eq.~(\ref{eqn8}) are degenerate, i.e. that successful outcome of
"$l_m$-fold elimination" of the state $|\gamma_m \rangle_c$,
corresponding to the root $\gamma_m$ with multiplicity $l_m$, (or,
in other words, successful elimination of $l_m$ "copies" of the
state $|\gamma_m \rangle_c$) corresponds to elimination of the
states $|\gamma_m \rangle_c$, $|\gamma_m^{(1)} \rangle_c$,~...,
$|\gamma_m^{(l_m-1)} \rangle_c$.

The successful result of "$l_m$-fold elimination" of the state
$|\gamma_m \rangle_c$ is described by operator
\begin{equation}
\label{eqn52} \hat A^{(n_1)}_ {|\gamma_m \rangle_c } ... \hat
A^{(n_{l_m})}_ {|\gamma_m \rangle_c } \propto \left( \hat c -
\gamma_m \right)^r,
\end{equation}
where $r=n_1+ ... + n_{l_m} \ge l_m$. The expression
\begin{equation*}
\label{eqn53} \left( \hat c - \gamma_m \right)^r \left( \hat c^+
\right)^s = \left( \hat c^+ + \frac{\partial}{\partial \hat c}
\right)^s \left( \hat c - \gamma_m \right)^r
\end{equation*}
contains powers of operator $\left( \hat c - \gamma_m \right)$ not
less than $r-s\ge l_m -s \ge 1$ for $s=0,... ,l_m - 1$. Therefore,
operator (\ref{eqn52}) corresponds to elimination of the states
$|\gamma_m^{(s)} \rangle$, $s=0,... l_m-1$:
\begin{equation}
\label{eqn54}\begin{aligned}& \left\{ \hat A^{(n_1)}_ {|\gamma_m
\rangle_c } ... A^{(n_{l_m})}_ {|\gamma_m \rangle_c } \right\} (\hat
c^+)^{s} |\gamma_m \rangle \propto {} \\ & {} \propto \left( \hat c
- \gamma_m \right)^{r-s} |\gamma_m \rangle =0\quad \mbox{for }
s=0,..., l_m-1,
\end{aligned}
\end{equation}
which proves the conclusion made in the previous subsection.

\subsection{Final state for successful and "semi-successful"
results of discrimination}

For obtaining the expression for the final state of the main field
modes $\hat a$ and $\hat b$ after discrimination of the state
$|\varphi \rangle_c$ by elimination measurements, it is convenient
to introduce "phase-shifting" operator $\hat F_{ab}=\exp\left( i
\chi \left(\hat a^+ \hat a+ \hat b^+ \hat b\right)\right)$ (it
describes change of the main field modes state after cross-Kerr
interaction with the mode $\hat c$ possessing 1 photon) and to
represent state $|\Psi_1 \rangle$ (described by Eq.~(\ref{eqn1}))
using operator of coherent displacement with operator-type argument:
\begin{equation}
\label{eqn55} |\Psi_1 \rangle_{abc} = \hat D_c \left( \hat F_{ab}
\gamma \right) |\alpha \rangle_a |\beta \rangle_b |0 \rangle_c,
\end{equation}
with the following property:
\begin{equation}
\label{eqn58} \hat c\, \hat D_c \left( \hat F_{ab} \gamma \right) =
\hat F_{ab} \gamma \hat D_c \left( \hat F_{ab} \gamma \right),
\end{equation}
which leads to significant simplification of Eq.~(\ref{eqn49}) for
the final state density matrix. As shown in Appendix~\ref{app1}
(Eqs.~(\ref{eqn59}), (\ref{eqn60})), the final state of main field
modes $\hat a$ and $\hat b$ after elimination of all the states
$\{|\gamma_j \rangle_c\}$ and subsequent discarding the ancillary
mode $\hat c$ is described by density matrix
\begin{equation}
\label{eqn56} \rho_{ab} = q^{2K} |\gamma|^{2K}
\left|{\Psi_f'}\mathrel{\left\rangle{\strut_{ab} {}
\strut_{ab}\vphantom{\Psi_f' }}\right\langle
\kern-\nulldelimiterspace}{\Psi_f'}\right| + O \left(
|\gamma|^{2K+2} \right),
\end{equation}
where
\begin{equation}
\label{eqn15} \begin{aligned} \left|{\Psi_f'}\right\rangle_{ab}  =
\left(\hat F_{ab}- \frac{\gamma_1}{\gamma} \right) ... \left(\hat
F_{ab}- \frac{\gamma_K}{\gamma} \right) \left|{\alpha}
\right\rangle_a \left|{\beta}\right\rangle_b =  \\
= \frac{1}{c_K}\sum\limits_{n=0} ^K c_n \hat F_{ab}^n \left|{\alpha
}\right\rangle_a \left|{\beta}\right\rangle_b  = \frac{1}{c_K}
\left| \Psi_f \right\rangle_{ab} .
\end{aligned}
\end{equation}
The obtained expression means that the distance between the desired
final state and the state $\rho_{ab}$, generated by the scheme, has
the order $O\left( |\gamma|^2 \right)$ and is small enough to be
neglected when nonideality of the system is taken into account.

It should be noted that in certain cases the final states, generated
when "clicks" were obtained not from all the detectors, can also be
useful ("semi-successful" results). Such states are described by
expressions, similar to Eq.~(\ref{eqn15}) but without multipliers,
corresponding to the detectors (with numbers $n_1$, $n_2$, ...)
that did not produce "clicks":
\begin{equation}
\label{eqn24}  \left|{\Psi(n_1, n_2, ...)}\right\rangle_{ab}  =
\prod_{m\ne n_1,n_2,...}\left(\hat F_{ab}- \frac{\gamma_m}{\gamma}
\right) \left|{\alpha }\right\rangle_a \left|{\beta}\right\rangle_b.
\end{equation}
This expression can be decomposed in the form, similar to the
desired final state (\ref{eqn3}) but with lower possible degree of
entanglement:
\begin{equation}
\label{eqn25}  \left|{\Psi(n_1, n_2,  ...)}\right\rangle_{ab} = \sum
\limits _{n}\tilde c_n (n_1, n_2,  ...) \left|{\alpha e^{i\chi
n}}\right\rangle_a \left|{\beta e^{i\chi n}}\right\rangle_b.
\end{equation}
E.g. for the case of absence of only one photocount the number of
terms equals to $K$ (instead of $K+1$) and the coefficients $\tilde
c_n$ can be found as
\begin{equation}
\label{eqn26} \tilde c_n (n_1) = \sum \limits_{m=0}^n \frac{c_m}{c_K
(\gamma_{n_1}/\gamma)^{n+1-m}},
\end{equation}
where $n=0,...,K-1$.

In the next section we provide several examples of final states that
can be generated by the protocol for successful and
"semi-successful" discrimination outcomes.

\section{Examples}
\label{sec:special}

\subsection{Superpositions with correlated photon numbers}

As the first example of possible applications of the protocol to
nonclassical states generation we consider creation of a
superposition of states of modes $\hat a$ and $\hat b$ with
correlated photon numbers. We show that such superpositions
arise quite naturally in our protocol and then use them to
illustrate general formalism, developed in
Section~\ref{sec:discrim}.

As discussed above (see Eq.~(\ref{eqn27})), cross-Kerr interaction
correlates photon number of one of the interacting modes with the
phase of coherent state of the other mode. The state
$|\Psi_1\rangle_{abc}$, obtained after cross-Kerr interaction of the
main modes $\hat a$ and $\hat b$ with the ancillary mode $\hat c$,
can be considered either as a superposition, where phases of
coherent states of the modes $\hat a$ and $\hat b$ are proportional
to the number of photons in the mode $\hat c$ (Eq. (\ref{eqn1})), or
alternatively as a superposition, where the phase of coherent state
of the mode $\hat c$ is proportional to the total number of photons
in the modes $\hat a$ and $\hat b$. The latter interpretation of the
state $|\Psi_1\rangle_{abc}$ implies that discrimination of coherent
state $|\gamma e^{i \chi n} \rangle_c$ of the ancillary mode $\hat
c$ fixes the total number of photons in modes $\hat a$ and $\hat b$
to be equal to $n$. The final state $|\Psi_f \rangle_{ab}$ in this
case is a superposition of Fock states of the modes $\hat a$ and
$\hat b$ with the number of photons in each mode varying form 0 to
$n$ and the total number of photons being equal to $n$ for each
term.

Mathematically this statement can be proved in the following way.
The state $|\Psi_1 \rangle_{abc}$ can be decomposed in the form:
\begin{equation}
\label{eqn34}  |\Psi_1 \rangle_{abc} =  \sum _{n=0} ^\infty |\Phi(n)
\rangle_{ab} |\gamma e^{i \chi n} \rangle_c,
\end{equation}
where
\begin{equation}
\label{eqn80} |\Phi(n) \rangle_{ab} = \sum _{m=0}^n Q_m (\alpha)
Q_{n-m} (\alpha) |m\rangle_a | n-m \rangle _b
\end{equation}
is a superposition of states of the modes $\hat a$ and $\hat b$ with
fixed total number of photons (equal to $n$); function $Q_m
(\alpha)$ is defined as $Q_m (\alpha) = \frac{\alpha^m}{\sqrt{m!}}
e^{-|\alpha|^2/2}$, $Q_m(\alpha) Q_{n-m} (\alpha) = Q_n (\alpha)
\sqrt{\binom{m}{n}}$; we assume for simplicity that $\alpha =
\beta$.

Suppose that only $K$ terms are significant in the superposition
(\ref{eqn34}): $|Q_n (\alpha) | \ll 1$ for $n>K$,
\begin{equation}
\label{eqn81} |\Psi_1 \rangle_{abc} \approx  \sum _{n=0} ^K |\Phi(n)
\rangle_{ab} |\gamma e^{i \chi n} \rangle_c.
\end{equation}
Then successful outcome of elimination of $K$ coherent states $\{
|\gamma \rangle_c, ..., |\gamma e^{i \chi (n-1) } \rangle_c, |\gamma
e^{i \chi (n+1) } \rangle_c, ..., |\gamma e^{i \chi K } \rangle_c
\}$ with subsequent discarding of the ancillary mode $\hat c$
transforms the state $|\Psi_1 \rangle _{abc}$ into the following
state of the modes $\hat a$ and $\hat b$ with correlated photon
numbers, described above:
\begin{equation}
\label{eqn105}\begin{aligned} |\Phi(n,K) \rangle_{ab} = 2^{-n/2}
\sum _{m=0}^n \sqrt{\binom{m}{n}} |m\rangle_a | n-m \rangle _b
\times{}
\\ \times \left\{ 1 + O \left( Q_{K+1}(\alpha) / Q_n(\alpha) \right)
\right\}.
\end{aligned}
\end{equation}

Generation of states of the form (\ref{eqn105}) can be described by
general formalism, developed in Section~\ref{sec:discrim}. For this
purpose we find coefficients $c_n$, for which the general final
state $|\Psi_f \rangle_{ab}$ (Eq.~(\ref{eqn3})) is equivalent for
the desired final state $|\Phi(n,K) \rangle_{ab}$
(Eq.~(\ref{eqn105})). Then coherent states amplitudes $\gamma_j$ can
be found by solving Eq.~(\ref{eqn8}), and expressions for the
parameters of discrimination scheme can be derived.

The general expression Eq.~(\ref{eqn3}) for the final state of the
modes $\hat a$ and $\hat b$ can be transformed to the following
form:
\begin{equation}
\label{eqn106} |\Psi_f \rangle_{ab} = \sum_{s=0}^\infty \left\{
\sum_{n=0}^K c_n e^{i \chi s n} \right\} |\Phi(s) \rangle_{ab},
\end{equation}
by decomposing coherent states $|\alpha e^{i \chi n} \rangle_a$ and
$|\beta e^{i \chi n} \rangle_b$ in terms of Fock states, where
states $|\Phi(s) \rangle_{ab}$ are defined by Eq.~(\ref{eqn80}).

In order to obtain $|\Psi_f \rangle_{ab} = |\Phi(s,K) \rangle_{ab}$,
the coefficients $c_n$ must satisfy the following system of
equations:
\begin{equation}
\label{eqn107} \sum_{n=0}^K c_n e^{i \chi s' n} = 0\; \mbox{for} \;
s'= 0,1,...,s-1,s+1,...,K.
\end{equation}
Before solving this system, it is useful to compare it with
Eq.~(\ref{eqn8}) for the amplitudes $\gamma_j$ and to notice, that
if coefficients $c_n$ satisfy Eq.~(\ref{eqn107}), $K$ complex
numbers $\gamma e^{i \chi s'}$, $s'= 0,1,...,s-1,s+1,...,K$,
apparently represent the $K$ roots of Eq.~(\ref{eqn8}). Then,
coefficients $c_n$ are defined in the unique way (except for overall
normalization constant) by the complete system of roots and are
equal to
\begin{equation}
\label{eqn108} c_0 = c_K \prod_{s'} e^{i \chi s'},\quad  ...,\quad
c_{K-1} =c_K \sum_{s'} e^{i \chi s'}.
\end{equation}

For example, if the desired final state is the following one
\begin{equation}
\label{eqn109}\begin{gathered} |\Psi_f\rangle_{ab} = |\Phi(2,2)
\rangle_{ab} = \\ {} = \frac{|0\rangle_a |2\rangle_b + \sqrt{2}
|1\rangle_a |1\rangle_b + |2\rangle_a |0\rangle_b}{2 \sqrt{2}} +
O\left( |\alpha| \right),
\end{gathered}
\end{equation}
coefficients $c_n$ must be equal to $c_0 = e^{ i \chi}$, $c_1 = -1
-e^{ i \chi}$, $c_2=1$ (for unnormalized state). Amplitudes of the
coherent states $|\gamma_j \rangle_c$, eliminated in discrimination
scheme, are equal to $\gamma_1 = \gamma$ and $\gamma_2 = \gamma e^{i
\chi}$ in this case. According to Eq.~(\ref{eqn35}), transmittances
of the beamsplitters $BS_1$ and $BS_2$ are equal approximately to
$T_1 \approx 1/2$ (for $\delta \ll 1$) and $T_2 = \delta$. The
amplitudes of the reference coherent states, defined by
Eq.~(\ref{eqn75}), are $\tilde \gamma_1 \approx - i \gamma_1 $ and
$\tilde \gamma_2 \approx - i \left(\gamma_1+
\gamma_2\right)/(\sqrt{2} \delta) $. Solving the system of equations
(\ref{eqn76}), one finds $\phi_1 = - \chi/2$, $T_1' \approx 1-
\delta^2 / 2$, $\tilde \gamma \approx - i \tilde \gamma_2$.

For coefficients $\tilde c_n(n_1)$ (Eq.~(\ref{eqn26})),
characterizing final state in the case of "semi-successful" outcomes
of discrimination when the desired final state for successful
outcome is described by Eq.~(\ref{eqn109}), one obtains the
following expressions: $\tilde c_0 (1) = e^{i \chi}$, $\tilde c_1
(1) = -1$, $\tilde c_0 (2) = 1$, $\tilde c_0 (2) = -1$. The final
state, generated when "click" was obtained from detector $D_2$, is
approximately a vacuum state:
\begin{equation}
\label{eqn110} |\Psi(1) \rangle_{ab} = |0 \rangle_a |0 \rangle_b +
O(|\alpha|^2),
\end{equation}
while the state, generated when "click" was obtained from detector
$D_1$, belongs to the class of states, described by
Eq.~(\ref{eqn105}):
\begin{equation}
\label{eqn111} |\Psi(2) \rangle_{ab} = \frac{|0 \rangle_a |1
\rangle_b + |1 \rangle_a |0 \rangle_b}{\sqrt{2}} + O(|\alpha|)
\equiv |\Phi(1,1)\rangle_{ab},
\end{equation}
and, therefore, can be useful for certain applications.

\subsection{Maximally entangled states for protocols with fixed number
of detectors}


Another group of examples represents states of the form
Eq.~(\ref{eqn3}) with maximal entanglement, which is possible for a
scheme with fixed number of photodetectors $K$ (and, therefore, with
fixed number of terms in the expression (\ref{eqn3}) for the final
state $|\Psi_f \rangle_{ab}$).

In the most simple case of schemes with 1 detector ($K=1$) the
coefficients $c_n$ in Eq.~(\ref{eqn3}), maximizing the final
entanglement, can be found analytically as
\begin{equation}
\label{eqn11}
\begin{gathered}
c_0 = 1,\\
c_1 =- \exp\left( - i \left( |\alpha| ^2 + |\beta| ^2 \right) \sin
\chi \right),
\end{gathered}
\end{equation}
where for simplicity we consider unnormalized final state.
Additional condition, required for maximization of entanglement in
this case, is $\left| \alpha \right|^2 = \left| \beta \right|^2 $
(for simplicity we will assume without loss of generality that
$\alpha = \beta$).

The final state of the system is approximately a Bell state
\begin{equation}
\label{eqn12} \left| \Psi^+ \right\rangle_{ab} = \left( |+\rangle_a
|- \rangle_b + |-\rangle_a |+ \rangle_b \right)/\sqrt 2,
\end{equation}
where $\left\{|+ \rangle_{a,b}, |- \rangle_{a,b} \right\}$ is the
orthonormal basis for the states of the modes $\hat a$ and $\hat b$:
$| \pm \rangle_{a,b} \sim ( |{\alpha}\rangle_{a,b} \pm e^{-i
|\alpha|^2 \sin \chi} |{\alpha e^{i \chi}}\rangle_{a,b} )$.
Therefore, the protocol can be used for generating qubit-type
quantum states with maximal entanglement (equal to 1), possible for
qubit systems.

The set of states $\left\{ | \gamma_m \rangle_c \right\}$ consists
of the only state $|\gamma_1 \rangle_c$ with the amplitude $\gamma_1
= \gamma e^{ i \left( |\alpha| ^2 + |\beta| ^2 \right) \sin \chi }$,
defined by Eq.~(\ref{eqn8}). The transmittance of the beam splitter
$BS_1$ (Fig.~\ref{fig2}(c)) equals $\delta$, and the amplitude of
the reference coherent state is $\tilde \gamma_1 = - i \gamma_1
\left(1-\delta\right) / \delta$.

For the scheme with two detectors ($K=2$) the coefficients $c_n$,
providing maximal entanglement, can be found analytically for
systems with $\chi \ll 1$ and $|\alpha| = |\beta|$ in two limiting
cases: $|\alpha|^2 \chi^2 \ll 1$ (low distinguishability of main
field modes coherent states with and without phase shift equal to
$\chi$: $ |\left\langle{\alpha e^{i
\chi}}\mathrel{\left|{\vphantom{\alpha e^{i \chi}}}\right.
\kern-\nulldelimiterspace}{\alpha}\right\rangle | \approx 1 $) and
$|\alpha|^2 \chi^2 \gg 1$ (high distinguishability: $ |
\left\langle{\alpha e^{i \chi}}\mathrel{\left|{\vphantom{\alpha e^{i
\chi}}}\right. \kern-\nulldelimiterspace}{\alpha}\right\rangle | \ll
1  $).

For $|\alpha|^2 \chi^2 \ll 1$ the final entanglement is maximal for
\begin{equation}
\label{eqn13}
\begin{gathered}
c_0 = 1,\\
c_1 = - 2 \left(1-|\alpha|^2 \chi^2 \right)
e^{-2 i |\alpha|^2 \chi}, \\
c_2 =  e^{-4 i |\alpha|^2 \chi} .
\end{gathered}
\end{equation}
The final state of the system is
\begin{equation}
\label{eqn14} \left|{\Psi_f}\right\rangle_{ab} = \frac{|u_1\rangle_a
|u_3\rangle_b + \sqrt{2} |u_2\rangle_a |u_2\rangle_b + |u_3\rangle_a
|u_1\rangle_b }{2},
\end{equation}
where
\begin{equation}
\label{eqn112} \begin{aligned} |{u_1}\rangle_a =  q_1 \bigl(
|{\alpha }\rangle_a e^{i |\alpha|^2 \chi} +  |{\alpha e^{2 i
\chi}}\rangle_a e^{-i |\alpha|^2 \chi} + {}\\ {} + q_0 |{\alpha e^{
i
\chi}}\rangle_a  \bigr), \\
|{u_2}\rangle_a = q_2 \bigl( |{\alpha }\rangle_a e^{i |\alpha|^2
\chi} - |{\alpha e^{2 i
\chi}}\rangle_a e^{-i |\alpha|^2 \chi}  \bigr), \\
|{u_3}\rangle_a =  q_3 \bigl( |{\alpha }\rangle_a e^{i |\alpha|^2
\chi} + |{\alpha e^{2 i \chi}}\rangle_a e^{-i |\alpha|^2 \chi} - {}\\
{} - q_0 |{\alpha e^{ i \chi}}\rangle_a \bigr),
\end{aligned}
\end{equation}
are orthonormal basis vectors for the mode $\hat a$ and the basis
vectors $|u_j\rangle_b$ for the mode $\hat b$ are defined in a
similar way (with $\alpha$ being replaced by $\beta$); coefficients
$q_i$ are determined by the condition of orthonormality of the
system of basis vectors. The final state (\ref{eqn14}) possesses
entanglement $E=3/2$, which is higher than the maximal value,
achievable for a pair of qubits.

In this case the coherent states $ | \gamma_1 \rangle_c $ and $ |
\gamma_2 \rangle_c $, exploited in the detection scheme, possess
close amplitudes $\gamma_{1,2} = \left( 1 \pm i \sqrt{2} |\alpha|
\chi-|\alpha|^2 \chi^2 \right) \gamma e^{ 2 i |\alpha| ^2 \chi }$.
Transmittances of the beamsplitters $BS_1$ and $BS_2$ are equal
approximately to $T_1 \approx 1/2$ (for $\delta \ll 1$) and $T_2 =
\delta$; amplitudes of the reference coherent states are $\tilde
\gamma_1 \approx - i \gamma_1 $ and $\tilde \gamma_2 \approx - i
\left(\gamma_1+ \gamma_2\right)/(\sqrt{2} \delta) $. The parameters
of discrimination scheme, defined by Eq.~(\ref{eqn76}), are equal to
$\phi_1 = \sqrt{2} |\alpha| \chi$, $T_1' \approx 1 - \delta^2 /2$,
$\tilde \gamma \approx - i \tilde \gamma_2$.

In the opposite limiting case $|\alpha|^2 \chi^2  \gg 1$ (such
condition is satisfied simultaneously with $\chi \ll 1$ in the
systems with intense fields $\hat a$ and $\hat b$) entanglement
reaches the bound for 3-level system $E=\log_2 3 \approx 1.58$ when
the parameters are
\begin{equation}
\label{eqn19}
\begin{gathered}
c_0 = 1, \\
c_1 = - e^{-2 i |\alpha|^2 \chi}, \\
c_2 =  e^{-4 i |\alpha|^2 \chi}.
\end{gathered}
\end{equation}

The final state in this case has the form
\begin{equation}
\label{eqn20} \left|{\Psi_f}\right\rangle_{ab} = \frac{|u_1\rangle_a
|u_3\rangle_b + |u_2\rangle_a |u_2\rangle_b + |u_3\rangle_a
|u_1\rangle_b }{\sqrt{3}},
\end{equation}
where basis vectors $|u_i \rangle_{a,b}$ are defined by
Eq.~(\ref{eqn112}) above. Amplitudes of coherent states $ | \gamma_1
\rangle $ and $ | \gamma_2 \rangle $ are equal to $\gamma_{1,2} =
\left(1 \pm i \sqrt{3}\right) e^{ 2 i |\alpha| ^2 \chi }/2  $. The
transmittances of the beamsplitters $BS_1$ and $BS_2$ and the
amplitudes $\tilde \gamma_1 $, $\tilde \gamma_2 $ of reference modes
coherent states are the same functions of the amplitudes $\gamma_1$
and $\gamma_2$ as in the previously discussed limiting case. Solving
the system of equations (\ref{eqn76}), one can find $\phi_1 = \pi
/3$, $T_1' \approx 1 - \delta^2 /2$, $\tilde \gamma \approx - i
\tilde \gamma_2$.

\begin{figure}[t]
\begin{center}
\includegraphics[scale=0.8]{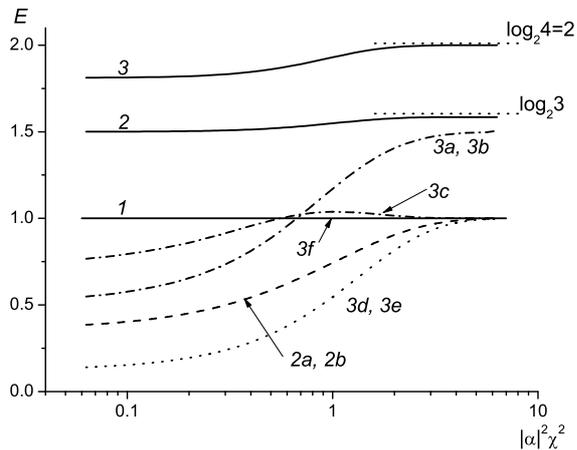}
\caption{Entanglement of the final states, generated by the
protocol, versus coherent states distinguishability $|\alpha|^2
\chi^2$. Solid lines: maximal entanglement, possible for generation
in the scheme with fixed number of photodetectors $K$: $K=1$ (1),
$K=2$ (2), $K=3$ (3). Dashed lines (2a, 2b): entanglement of the
final state, generated by the scheme with $K=2$ detectors with
optimized coefficients $c_n$ in "semi-successful" case (one
photocount instead of two; entanglement does not depend on the
number of the detector, which "clicked"). Dot-dashed lines (3a, 3b,
3c): final state entanglement for $K=3$ and 1 absent photocount.
Dotted lines (3d, 3e, 3f; line 3f coincides with the solid line 1):
final entanglement for $K=3$ and 2 absent photocounts.} \label{fig3}
\end{center}
\end{figure}

For intermediate values of distinguishability ($|\alpha|^2 \chi^2
\sim 1$), as well as for greater numbers of detectors ($K\ge3$),
optimal coefficients $c_n$ can be found numerically. The values of
maximally possible entanglement for schemes with fixed number of
detectors are shown in Fig.~\ref{fig3} (solid lines). One can see
that the maximal possible value of entanglement grows with increase
of the detectors number $K$, and in certain cases it may be
considered as more effective use of fixed resources (nonlinear
interaction, quantum channel) than can be achieved in schemes with
qubit-type entanglement.

Fig.~\ref{fig3} also illustrates entanglement of final states,
generated by these schemes for "semi-successful" outcomes (obtaining
photocounts from lesser number of detectors; dashed, dotted and
dot-dashed lines in Fig.~\ref{fig3}). These final states possess
non-zero entanglement and in certain cases can also be useful for
solving information processing tasks. For example, the scheme with
$K=3$ detectors, optimized for obtaining maximal entanglement in the
case of successful discrimination outcome, in the case of two absent
photocounts can also produce maximally entangled state from the
space of states $|\Psi_f \rangle_{ab}$ with 2 non-zero terms (line
3f in Fig.~\ref{fig3}).

\section{Nonideal system}

\subsection{Considered types of nonideality}

In previous sections we assumed for simplicity that the quantum
channel and photodetectors are ideal. In order to prove
realizability of the proposed entanglement generation method in
realistic situations we discuss influence of system nonideality
on the fidelity of obtaining final state.

In real system decoherence and dephasing accompany all the stages of
the protocol: implementation of nonlinear interaction, transmission
of the probe beam through the quantum channel, storage of main modes
$\hat a$ and $\hat b$ in local resonators. All these factors can be
taken into account by solving corresponding master equations
\cite{mogilevtsev_korolkova}. However, for the system, considered in
our paper, influence of some of the factors on the fidelity of final
state generation is supposed to be small. Therefore, for the purpose
of simplifying further description, we take into account only
the following factors that can limit applicability of our protocol:
\renewcommand\labelenumi{\theenumi}

\renewcommand\theenumi{(\roman{enumi})}
\begin{enumerate}
\item\label{nonid1} decoherence of modes $\hat a$ and $\hat b$ during
cross-Kerr interaction with probe beam $\hat c$ (decoherence of the
mode $\hat c$ is assumed to have negligible effect due to much
smaller amplitude of the probe beam: $|\gamma| \ll |\alpha|,
|\beta|$; qualitatively, decoherence of the mode $\hat c$ during
cross-Kerr interaction influences the final state in the same way as
decoherence of this mode during transmission through the quantum
channel, but is weaker);
\item \label{nonid2} inaccuracies in the nonlinearity values of the used
Kerr media (we assume that effective strengthes of nonlinear
interactions carried out by Alice and Bob are equal to $\chi _{ac} =
\chi + \Delta \chi_{ac}$ and $\chi _{bc} = \chi + \Delta \chi_{bc}$
respectively and differ from the value $\chi$ used in the scheme
optimization);
\item \label{nonid3} decoherence caused by nonideality of local
resonator at Alice's site (Bob's resonator is not used for storing
part of an entangled state for a long time, and its nonideality is
supposed to effect the final state fidelity negligibly; dephasing in
the resonators is assumed to be small due to controllable laboratory
conditions);
\item \label{nonid4} decoherence and dephasing in the quantum channel;
\item \label{nonid5} limited efficiency (probability of detecting a
photon, present in the field mode, equals $\lambda < 1$) and dark
counts of photodetectors (obtaining photocount with probability
$\zeta$ when the mode is in the vacuum state).
\end{enumerate}

For describing nonideality of the system we consider four
stages of the protocol separately and find (super-)operators,
describing difference between the states, obtained in ideal and
nonideal systems. It should be noted that operator, transforming one
fixed state into another fixed state (ideal state into nonideal one
in the considered case), is not defined in the unique way: its
action on states from orthogonal space can be arbitrary. In further
consideration we try to choose operators, acting on the main
modes only (but not on the ancillary one), from sets of equivalent
operators, transforming ideal density matrix into the nonideal one.
Then the final state after implementing all the stages of the
protocol is expected to be presented in the form of certain
superoperator, acting on the ideal final state $|\Psi_f\rangle_{ab}$
of the main modes $ \hat a$ and $\hat b$.

\subsection{Nonideal cross-Kerr interaction at Alice's site}

The first stage of the protocol is cross-Kerr interaction of modes
$\hat a$ and $\hat c$. We describe this nonlinear interaction
in nonideal case by the following master equation:
\begin{equation}
\label{eqn61} \frac{d}{dt} \rho = i \chi' \left[ \hat a^+ \hat a
\hat c^+ \hat c , \rho \right] + \kappa_1 \hat L (\hat a ) \rho,
\end{equation}
where $\hat L(\hat X) \rho = \left[ \hat X \rho, \hat X \right] +
\left[ \hat X , \rho \hat X \right]$. As discussed above (item
\ref{nonid1} in the list of nonideality types), the only kind of
nonideality taken into account by Eq.~(\ref{eqn61}) is decoherence
of the mode $\hat a$.

Assuming that the duration of interaction is $\Delta t_1$, one can
characterize the nonlinear interaction by effective strength
$\chi_{ac} = \chi' \Delta t_1$ and relative loses rate $\Lambda _1 =
e^{2 \kappa_1 \Delta t_1} - 1$ ($\Lambda_1 = (I_0-I)/I$, where $I_0$
and $I$ are beam intensities before and after the interaction).

In the ideal case ($\kappa_1 = 0$) state transformation due to
discussed cross-Kerr interaction is described by unitary operator
\begin{equation}
\label{eqn113} \hat U_{ac} = \exp(i \chi_{ac} \hat a^+ \hat a \hat
c^+ \hat c) .
\end{equation}
The initial uncorrelated state $|\alpha\rangle_a |\beta\rangle_b
|\gamma\rangle_c$ is transformed, therefore, into the superposition,
where the phase of coherent state of the mode $\hat a$ is
proportional to the number of photons in the mode $\hat c$:
\begin{equation}
\label{eqn114} \hat U_{ac}  |\alpha\rangle_a |\beta\rangle_b
|\gamma\rangle_c = \sum_{n=0}^\infty Q_n(\gamma) |\alpha e^{i
\chi_{ac} n} \rangle_a |\beta\rangle_b |n\rangle_c.
\end{equation}

Transformation of the system state in the nonideal case can be found
by solving the master equation Eq.~(\ref{eqn61}) with the initial
condition $\rho(0) = \rho_{abc}^{(0)}$, where
\begin{equation} \label{eqn64} \rho_{abc}^{(0)} =
|{\alpha^{(0)}}\mathrel{\rangle \strut_a \strut_a \langle
\kern-\nulldelimiterspace}{\alpha^{(0)}}| \otimes
|{\beta^{(0)}}\mathrel{\rangle \strut_b \strut_b \langle
\kern-\nulldelimiterspace}{\beta^{(0)}}| \otimes
|{\gamma^{(0)}}\mathrel{\rangle \strut_c \strut_c \langle
\kern-\nulldelimiterspace}{\gamma^{(0)}}|
\end{equation}
is the initial uncorrelated density matrix (initial amplitudes
$\alpha^{(0)}$, $\beta^{(0)}$, $\gamma^{(0)}$ of coherent states in
the nonideal case must be larger that the amplitudes $\alpha$,
$\beta$, $\gamma$ that are expected to characterize final state).

The solution can be found by representing the density matrix of the
modes $\hat a$ and $\hat c$ in the form
\begin{equation}
\label{eqn17}\begin{aligned} & \rho(t)= \sum_{n_1,n_2} \rho_{n_1
n_2}(t) \times {} \\ &{} \times \left|{\alpha(t)  e^{i \chi' t
n_1}}\mathrel{\left\rangle \strut_a \strut_a {\vphantom{\alpha(t)
e^{i \chi' t n_1} \alpha(t)}}\right\langle
\kern-\nulldelimiterspace}{\alpha(t)
 e^{i
\chi' t n_2}}\right| \otimes \left|{n_1}\mathrel{\left\rangle
\strut_c \strut_c {\vphantom{n_1 n_2}}\right\langle
\kern-\nulldelimiterspace}{n_2}\right|,
\end{aligned}
\end{equation}
which is preserved during evolution. Substituting this decomposition
into Eq.~(\ref{eqn61}) and solving the resulting system of
differential equations, one can obtain the following expressions for
the quantities $\alpha(\Delta t_1)$ and $\rho_{n_1 n_2} (\Delta
t_1)$ at the end of the considered stage of the protocol:
\begin{equation}
\label{eqn18} \alpha(\Delta t_1) = \alpha ' \equiv \alpha^{(0)} e^{-
\kappa_1 \Delta t_1} \equiv \alpha^{(0)} / \sqrt{\Lambda_1 + 1 },
\end{equation}
\begin{equation}
\label{eqn89}\begin{aligned} \rho_{n_1 n_2}(\Delta t_1) \approx
Q_{n_1} (\gamma^{(0)}) Q_{n_2}^\ast  (\gamma^{(0)}) \exp
\bigl\{|\alpha^{(0)}|^2 \times{} \\ {} \times \Lambda_1 \bigl( i
\chi_{ac}(n_1 - n_2) - \chi_{ac}^2 (n_1 - n_2)^2 \bigr) \bigr\},
\end{aligned}
\end{equation}
where smallness of nonideality is assumed for simplicity of derived
expressions. The exponential factor in Eq.~(\ref{eqn89}) describes
influence of Kerr medium nonideality on the state, obtained after
the interaction. For characterizing transition from the ideal state
Eq.~(\ref{eqn114}) to the nonideal one the following superoperator,
acting in the state space of the mode $\hat a$, can be chosen from
the class of equivalent operators, describing this state
transformation:
\begin{equation}
\label{eqn115} \begin{aligned}  &\hat M_1 (\hat a) \colon
\bigl|{\alpha^{(0)} e^{i \chi_{ac} n_1}}\mathrel{\bigl\rangle{
\strut_a \strut_a \vphantom{\alpha^{(0)} e^{i \chi_{ac} n_1}
\alpha^{(0)} e^{i \chi_{ac} n_2}}}\bigr\langle
\kern-\nulldelimiterspace}{\alpha^{(0)} e^{i \chi_{ac} n_2}}\bigr|
\mapsto {} \\ & {} \mapsto \bigl|{\alpha' e^{i \chi_{ac}
n_1}}\mathrel{\bigl\rangle{ \strut_a \strut_a \vphantom{\alpha' e^{i
\chi_{ac} n_1} \alpha' e^{i \chi_{ac} n_2}}}\bigr\langle
\kern-\nulldelimiterspace}{\alpha' e^{i \chi_{ac} n_2}}\bigr| \cdot
\exp \Bigl\{|\alpha^{(0)}|^2 \times{} \\ &{} \times \Lambda_1
\Bigl(\frac{1}{2} i \chi_{ac}(n_1 - n_2) - \frac{1}{3}\chi_{ac}^2
(n_1 - n_2)^2 \Bigr) \Bigr\}.
\end{aligned}
\end{equation}
This superoperator adds small phase shift to the coefficients before
coherent states of the mode $\hat a$ and decreases non-diagonal
elements of the density matrix.

Another factor, which influences fidelity of the final state
generation but is not connected with nonideality of the Kerr medium
itself, is deviation of the nonlinearity effective strength
$\chi_{ac}$ from its expected value $\chi$, used during optimization
of the discrimination scheme parameters. This factor can be
accounted for by introducing superoperator
\begin{equation}
\label{eqn70a} \begin{aligned}  \hat M_\chi (\hat a) \colon {}&
\bigl|{\alpha^{(0)} e^{i \chi n_1}}\mathrel{\bigl\rangle{ \strut_a
\strut_a \vphantom{\alpha^{(0)} e^{i \chi n_1} \alpha^{(0)} e^{i
\chi n_2}}}\bigr\langle \kern-\nulldelimiterspace}{\alpha^{(0)} e^{i
\chi n_2}}\bigr| \mapsto {} \\ & {} \mapsto \bigl|{\alpha^{(0)} e^{i
\chi_{ac} n_1}}\mathrel{\bigl\rangle{ \strut_a \strut_a
\vphantom{\alpha^{(0)} e^{i \chi_{ac} n_1} \alpha^{(0)} e^{i
\chi_{ac} n_2}}}\bigr\langle \kern-\nulldelimiterspace}{\alpha^{(0)}
e^{i \chi_{ac} n_2}}\bigr|,
\end{aligned}
\end{equation}
which provides additional phase shift to coherent states of the mode
$\hat a$.

\subsection{Storage of the mode $\hat a$ in nonideal local
resonator at Alice's site}

The second stage of the protocol consists in transmission of the
ancillary field from Alice to Bob. At the same time the mode $\hat
a$, already correlated with the ancillary mode $\hat c$, is stored
at Alice's site. These two modes interact with the environment
independently, and corresponding kinds of nonideality are
considered separately.

Decoherence of the mode $\hat a$ in nonideal local resonator is
described by the following master equation:
\begin{equation}
\label{eqn116} \frac{d}{dt} \rho = \kappa_2 \hat L (\hat a ) \rho.
\end{equation}
If duration of this stage is equal to $\Delta t_2$, relative losses
rate equals to $\Lambda _2 = e^{2 \kappa_2 \Delta t_2} - 1$.

The solution of Eq.~(\ref{eqn116}) can be found in the way, similar to
the one used for the previous stage of the protocol. The amplitude of
coherent states of the mode $\hat a$ $\alpha(\Delta t_2) = \alpha '
e^{- \kappa_2 t_2}$ after this stage of the protocol must be equal
to its final value $\alpha$.

The influence of decoherence in Alice's local resonator on the state of
the system can be described by the superoperator
\begin{equation}
\label{eqn117} \begin{aligned}  &\hat M_2 (\hat a) \colon
\bigl|{\alpha' e^{i \chi_{ac} n_1}}\mathrel{\bigl\rangle{ \strut_a
\strut_a \vphantom{\alpha' e^{i \chi_{ac} n_1} \alpha' e^{i
\chi_{ac} n_2}}}\bigr\langle \kern-\nulldelimiterspace}{\alpha' e^{i
\chi_{ac} n_2}}\bigr| \mapsto {} \\ & {} \mapsto \bigl|{\alpha e^{i
\chi_{ac} n_1}}\mathrel{\bigl\rangle{ \strut_a \strut_a
\vphantom{\alpha e^{i \chi_{ac} n_1} \alpha e^{i \chi_{ac}
n_2}}}\bigr\langle \kern-\nulldelimiterspace}{\alpha e^{i \chi_{ac}
n_2}}\bigr| \cdot \exp \Bigl\{|\alpha'|^2 \times{} \\ &{} \times
\Lambda_2 \Bigl( i \chi_{ac}(n_1 - n_2) - \frac{1}{2}\chi_{ac}^2
(n_1 - n_2)^2 \Bigr) \Bigr\},
\end{aligned}
\end{equation}
transforming the system state in the same way as $\hat M_1(\hat a)$.

\subsection{Transmission of the ancillary field through nonideal
quantum channel}

Interaction with the environment of the ancillary mode $\hat c$
during the second stage of the protocol is described by master
equation
\begin{equation}
\label{eqn62} \frac{d}{dt} \rho = \kappa \hat L (\hat c ) \rho +
\Gamma \hat L (\hat c^+ \hat c) \rho ,
\end{equation}
where the first and the second terms describe decoherence and
dephasing of the mode $\hat c$ respectively (see item \ref{nonid4}
in the list of nonideality factors). For characterizing this type of
system nonideality one can introduce relative losses rate $\Lambda =
e^{2 \kappa \Delta t_2} - 1$ and mean phase error $\Delta \phi =
\sqrt{\Gamma \Delta t_2 }$.

Due to commutativity of $\hat L(\hat c) $ and $\hat L (\hat c^+ \hat
c)$ (in the sense that $\hat L(\hat c) \hat L (\hat c^+ \hat c) \rho
= \hat L (\hat c^+ \hat c) \hat L(\hat c) \rho $), the master
equation Eq.~(\ref{eqn62}) can be divided into two independent
parts, describing decoherence and dephasing.

State transformation because of decoherence has the form:
\begin{equation}
\label{eqn65} \rho \mapsto \sum_{n=0}^\infty \frac{(1-e^{-2 \kappa
\Delta t_2})^n}{n!}\, \hat c^n \rho  (\hat c^+)^n.
\end{equation}
Dephasing of the mode $\hat c$ transforms the system state as
\begin{equation}
\label{eqn69} |{n_1}\mathrel{\rangle{ \strut_c \strut_c
\vphantom{n_1 n_2}}\langle \kern-\nulldelimiterspace}{n_2}| \mapsto
|{n_1}\mathrel{\rangle{ \strut_c \strut_c \vphantom{n_1 n_2}}\langle
\kern-\nulldelimiterspace}{n_2}| \cdot e^{-\Delta \phi^2 (n_1 -
n_2)^2}.
\end{equation}

As stated above, for certain simplification of further consideration
it is useful to choose superoperators, acting on the main field
modes only, from the set of equivalent superoperators, describing
transition between ideal and nonideal case. For this purpose we take
into account that in the expression (\ref{eqn114}) for the ideal
system state number states $|n \rangle_c$ of the ancillary mode
appear in pairs with coherent states $|\alpha e^{i \chi n}
\rangle_a$. Therefore, decrease of non-diagonal density matrix
elements in the basis of Fock states of the mode $\hat c$ is
equivalent to corresponding decrease of non-diagonal elements for
the mode $\hat a$ in the basis of coherent states. In a similar way,
discrete changes of photon number in the mode $\hat c$, caused by
energy losses in the quantum channel and described by
Eq.~(\ref{eqn65}), are equivalent to corresponding discrete changes
of phase of the mode $\hat a$. After certain mathematical
calculations, one can show that the difference between ideal and
nonideal states, caused by decoherence and dephasing of the mode
$\hat c$, can be described by superoperators
\begin{equation}
\label{eqn71} \hat M_2^{(1)}(\hat a) \colon \rho \mapsto
\sum_{n=0}^\infty \frac{(\Lambda |\gamma|^2)^n}{n!} e^{i \chi_{ac} n
\hat a^+ \hat a } \rho  e^{-i \chi_{ac} n \hat a^+ \hat a }.
\end{equation}
and
\begin{equation}
\label{eqn78}\begin{aligned} \hat M_2^{(2)} &(\hat a) \colon
\left|{\alpha e^{i \chi_{ac} n_1}}\mathrel{ \left\rangle{ \strut_a
\strut_a \vphantom{\alpha e^{i \chi_{ac} n_1} \alpha e^{i \chi_{ac}
n_2}}}\right\langle \kern-\nulldelimiterspace}{\alpha e^{i \chi_{ac}
n_2}}\right| \mapsto {} \\& {} \mapsto \left|{\alpha e^{i \chi_{ac}
n_1}}\mathrel{\left\rangle{ \strut_a \strut_a \vphantom{\alpha e^{i
\chi_{ac} n_1} \alpha e^{i \chi_{ac} n_2}}}\right\langle
\kern-\nulldelimiterspace}{\alpha e^{i \chi_{ac} n_2}}\right|
e^{-\Delta \phi^2 (n_1 - n_2)^2}.
\end{aligned}
\end{equation}
respectively.

\subsection{Nonideal cross-Kerr interaction at Bob's site}

The third stage of the protocol is cross-Kerr interaction of modes
$\hat b$ and $\hat c$, described by the following master equation:
\begin{equation}
\label{eqn63} \frac{d}{dt} \rho = i \chi'' \left[ \hat b^+ \hat
b\hat c^+ \hat c , \rho \right] + \kappa_3 \hat L (\hat b ) \rho,
\end{equation}
and characterized by effective nonlinearity strength $\chi_{bc} =
\chi '' \Delta t_3$ and relative losses rate $\Lambda_3 = e^{2
\kappa_3 \Delta t_3} - 1$ (we will assume for simplicity that
$\Lambda_3 = \Lambda_1$), where $\Delta t_3$ is the duration of the
interaction.

In the ideal case this nonlinear interaction is described by
operator
\begin{equation}
\label{eqn68} \hat U_{bc} = \exp(i \chi_{bc} \hat b^+ \hat b \hat
c^+ \hat c) ,
\end{equation}
with the obtained state $\hat U_{bc} \hat U_{ac}
|\alpha\rangle_a |\beta\rangle_b |\gamma\rangle_c$ being equal to
$|\Psi_1\rangle_{abc}$ (see Eq.~(\ref{eqn1})).

For nonideal system, master equation Eq.~(\ref{eqn63}) can be solved
exactly in the same way as Eq.~(\ref{eqn61}). The superoperator,
describing transition between ideal and nonideal cases, has the form
\begin{equation}
\label{eqn66}  \begin{aligned}  &\hat M_3 (\hat b) \colon
\bigl|{\beta^{(0)} e^{i \chi_{bc} n_1}}\mathrel{\bigl\rangle{
\strut_b \strut_b \vphantom{\beta^{(0)} e^{i \chi_{bc} n_1}
\beta^{(0)} e^{i \chi_{bc} n_2}}}\bigr\langle
\kern-\nulldelimiterspace}{\beta^{(0)} e^{i \chi_{bc} n_2}}\bigr|
\mapsto {} \\ & {} \mapsto \bigl|{\beta e^{i \chi_{bc}
n_1}}\mathrel{\bigl\rangle{ \strut_b \strut_b \vphantom{\beta e^{i
\chi_{bc} n_1} \beta e^{i \chi_{bc} n_2}}}\bigr\langle
\kern-\nulldelimiterspace}{\beta e^{i \chi_{bc} n_2}}\bigr| \cdot
\exp \Bigl\{|\beta^{(0)}|^2 \times{}
\\ &{} \times \Lambda_1 \Bigl(\frac{1}{2} i \chi_{bc}(n_1 - n_2) -
\frac{1}{3}\chi_{bc}^2 (n_1 - n_2)^2 \Bigr) \Bigr\}
\end{aligned}
\end{equation}
and, similarly to $\hat M_1 (\hat a)$, adds small phase shift to the
coefficients before coherent states of the mode $\hat b$ and
decreases non-diagonal elements of the density matrix.

Deviation of the nonlinearity effective strength $\chi_{bc}$ from
its expected value $\chi$ can be accounted for by introducing the
following superoperator, describing additional phase shift to
coherent states of the mode $\hat b$:
\begin{equation}
\label{eqn70b} \begin{aligned}  \hat M_\chi (\hat b) \colon {}&
\bigl|{\beta^{(0)} e^{i \chi n_1}}\mathrel{\bigl\rangle{ \strut_b
\strut_b \vphantom{\beta^{(0)} e^{i \chi n_1} \beta^{(0)} e^{i \chi
n_2}}}\bigr\langle \kern-\nulldelimiterspace}{\beta^{(0)} e^{i \chi
n_2}}\bigr| \mapsto {} \\ & {} \mapsto \bigl|{\beta^{(0)} e^{i
\chi_{bc} n_1}}\mathrel{\bigl\rangle{ \strut_b \strut_b
\vphantom{\beta^{(0)} e^{i \chi_{bc} n_1} \beta^{(0)} e^{i \chi_{bc}
n_2}}}\bigr\langle \kern-\nulldelimiterspace}{\beta^{(0)} e^{i
\chi_{bc} n_2}}\bigr|.
\end{aligned}
\end{equation}

\subsection{Nonideality of discrimination scheme due to limited efficiency
and dark counts of photodetectors}

The last stage of the protocol is discrimination of the state
$|\varphi \rangle_c$ of the mode $\hat c$ at Bob's site. This stage
includes operations on the ancillary mode $\hat c$ only and does not
influence directly modes $\hat a$ and $\hat b$ (their state is
transformed due to previously generated correlations with the
ancillary mode). All the superoperators $\hat M_1 (\hat a)$, $\hat
M_\chi (\hat a)$, $\hat M_2 (\hat a)$, $\hat M_2^{(1)} (\hat a)$,
$\hat M_2 ^{(2)} (\hat a)$, $\hat M_3 (\hat b)$, $\hat M_\chi (\hat
b)$, introduced for describing nonideality of the three preceding
stages of the protocol, act on the state spaces of the mode $\hat a$
and $\hat b$. Therefore, they must commute with any superoperators,
characterizing the last stage of the protocol, and can be considered
as acting \emph{after} implementation of nonideal discrimination
measurement. For such consideration the input state of the
discrimination scheme is the state $|\Psi_1 \rangle_{abc}$, defined
by Eq.~(\ref{eqn1}).

For the ideal system, obtaining successful outcome of all the
elimination measurements, followed by discarding of the mode $\hat
c$, transforms the input weakly entangled state
$|\Psi_1\rangle_{abc}$ into the desired final state $|\Psi_f
\rangle_{ab}$ (see Eqs.~(\ref{eqn56}), (\ref{eqn15})). Probability
of this successful outcome equals $p_K^{(ideal)} =  \left( q
|\gamma|^2 \right)^K /{|c_K|^2}$.

Limited efficiency of detectors (described by the probability
$\lambda$ of registering photons --- see item \ref{nonid5} in the
list of types of system nonideality) leads to decrease of
probability of obtaining successful discrimination outcome,
effectively reducing fraction $q$ of the coherent state amplitude of
the ancillary mode, interacting with photodetectors, by factor
$\lambda$: $q \mapsto \lambda q$. Then one-run success probability
for nonideal detection scheme is equal to
\begin{equation}
\label{eqn82} p_K =  \left( q \lambda |\gamma|^2 \right)^K
\frac{1}{|c_K|^2} \approx \left( \frac{\lambda |\gamma|^2}{K}
\right)^K \frac{1}{|c_K|^2}.
\end{equation}

Dark counts of photodetectors lead to mixing density matrices,
characteristic to "semi-successful" outcomes, to the final density
matrix, corresponding to successful elimination of all the states
$|\gamma_j \rangle_c$. Then, according to
Eqs.~(\ref{eqn56})--(\ref{eqn26}), successful discrimination of the
state $|\varphi \rangle_c$ by the scheme with nonideal
photodetectors transforms the state $|\Psi_1 \rangle_{abc} $ into
the following mixed state:
\begin{equation}
\label{eqn97}\begin{aligned} &\rho_{ab}' = |\Psi_f \mathrel{ \rangle
\strut_{ab} \strut_{ab} \langle} \Psi_f | +{}
\\ &{} + \frac{ \zeta}{\lambda |\gamma|^2} |c_K|^2 \sum_{n_1} |\Psi(n_1)
\mathrel{\rangle \strut_{ab} \strut_{ab} \langle} \Psi(n_1) | +{} \\
&{} + \frac{ \zeta^2}{\lambda^2 |\gamma|^4} |c_K|^2 \sum_{n_1,n_2}
|\Psi(n_1,n_2) \mathrel{\rangle \strut_{ab} \strut_{ab} \langle}
\Psi(n_1,n_2) |+{} \\
&{} + ...,
\end{aligned}
\end{equation}
where $j$-th term corresponds to presence of $j-1$ dark counts.

\subsection{Final state in the nonideal case}

Summarizing the results concerning discussed types of system
nonideality, we can express the density matrix of the final state of
the modes $\hat a$ and $\hat b$ in the following form:
\begin{equation}
\label{eqn77}\begin{aligned} \rho_{ab}^{(final)} = \hat M_3 (\hat b)
\hat M_{\chi}(\hat b)  \hat M_2^{(2)} (\hat a) \hat M_2^{(1)} (\hat
a) \hat M_2 (\hat a) \times {} \\ {} \times \hat M_1 (\hat a) \hat
M_\chi (\hat a) \rho_{ab}'.
\end{aligned}
\end{equation}

This expression can be simplified by taking into account that
coherent states of the modes $\hat a$ and $\hat b$ posses correlated
phases in all terms of the expression for the density matrix
$\rho_{ab}'$ (Eq.~(\ref{eqn97})) and appear only in groups of the
form $c_n |\alpha e^{i \chi n} \rangle_a |\beta e^{ i\chi n}
\rangle_b$ or $\tilde c_n(...) |\alpha e^{i \chi n} \rangle_a |\beta
e^{ i\chi n} \rangle_b$. Therefore, superoperators $\hat M_1 (\hat
a)$, $\hat M_2 (\hat a)$, $\hat M_2^{(2)} (\hat a)$, $\hat M_3 (\hat
b)$ act at the system state in the same way: they add small phase
shifts to the coefficients $c_n$ (or $\tilde c_n(...)$) and
decreases non-diagonal elements of the density matrix. Due to
commutativity of the superoperators, they can be collected in a
single superoperator
\begin{equation}
\label{eqn22} \hat M_0 (\hat a, \hat b) = \hat M_3 (\hat b) \hat
M_2^{(2)} (\hat a)  \hat M_2 (\hat a) \hat M_1 (\hat a),
\end{equation}
transforming pairs of coefficients $c_{n_1} c_{n_2}^\ast$ of the
state $|\Psi_f \mathrel{\rangle \langle} \Psi_f |$ (as well as pairs
of coefficients $\tilde c_n(n_1,n_2,...)$ with the similar meaning,
defined by Eq.~(\ref{eqn26})) in the following way:
\begin{equation}
\label{eqn85} c_{n_1} c_{n_2}^\ast \mapsto c_{n_1} c_{n_2}^\ast e^ {
i \eta_1 (n_1 - n_2) - \eta _ 2 (n_1 - n_2)^2 },
\end{equation}
where
\begin{equation}
\label{eqn86} \eta_1 = \frac{1}{2} \Lambda_1 \left( |\alpha|^2
\chi_{ac}  +  |\beta|^2 \chi_{bc}\right) + |\alpha|^2 \chi_{ac}
\Lambda_2
\end{equation}
is a phase difference per photon and
\begin{equation}
\label{eqn87} \eta_2 = \Delta \phi^2 + \frac{1}{3} \Lambda_1 \left(
|\alpha|^2 \chi_{ac}^2  +  |\beta|^2 \chi_{bc}^2\right) +
\frac{1}{2}|\alpha|^2 \chi_{ac}^2 \Lambda_2
\end{equation}
describes decay of non-diagonal elements of density matrix.

The first term in the exponent of Eq.~(\ref{eqn85}) corresponds to
changing phase of coefficients $c_n$ and can be compensated by
corresponding changes in the detection scheme (by replacing
$\gamma_j$ by $\gamma_j e^{-i\eta_1}$ in expressions for the scheme
parameters). Therefore, only the second term of the exponent is
essential for estimation of the deviation of the nonideal final
state from the ideal one.

Finally, expression (\ref{eqn77}) for the density matrix, obtained
in for nonideal system, can be rewritten using the notations of
Eq.~(\ref{eqn22}) in the form
\begin{equation}
\label{eqn79} \rho_{ab}^{(final)} = \hat M_0 (\hat a, \hat b) \hat
M_2^{(1)} (\hat a) \hat M_{\chi}(\hat a, \hat b) \rho_{ab}',
\end{equation}
where $\hat M_{\chi}(\hat a, \hat b) = \hat M_{\chi}(\hat b) \hat
M_{\chi}(\hat a)$.

In further consideration we assume that all types of
nonideality, present in the system, are weak enough and the fidelity
is close to unity (such systems are most useful from the practical
point of view). Then the density matrix, defined by
Eq.~(\ref{eqn79}), is approximately equal to
\begin{equation}
\label{eqn74} \begin{aligned} \rho_{ab}^{(final)} \approx {}&
\left|{\Psi_f}\mathrel{\left\rangle{ \strut_{ab} \strut_{ab}
\vphantom{\Psi_f \Psi_f}}\right\langle
\kern-\nulldelimiterspace}{\Psi_f}\right| + {} \\ {} + {} & \Bigl\{
\Delta \hat M_0 (\hat a, \hat b)  + \Delta\hat M_2^{(1)} (\hat a) +
{} \\ &{} + \Delta \hat M_\chi (\hat a, \hat b) \Bigr\}
\left|{\Psi_f}\mathrel{\left\rangle{ \strut_{ab} \strut_{ab}
\vphantom{\Psi_f \Psi_f}}\right\langle
\kern-\nulldelimiterspace}{\Psi_f}\right|  + {} \\ {} + {} & \frac{
\zeta}{\lambda |\gamma|^2} |c_K|^2 \sum_{n_1} |\Psi(n_1)
\mathrel{\rangle \strut_{ab} \strut_{ab} \langle} \Psi(n_1) |,
\end{aligned}
\end{equation}
where notation $\Delta \hat M_i = \hat I - \hat M_i$ is introduced
for the superoperators; only the leading order of small parameters,
characterizing nonideality of the system, is taken into account.

\subsection{Fidelity of the final state generation}

According to the standard definition, the fidelity of generating the
desired final state $|\Psi_f \rangle_{ab}$ equals
\begin{equation}
\label{eqn83} F = \frac{
\strut_{ab}\langle{\Psi_f}\mathrel{|{\vphantom{\Psi_f}
\rho_{ab}^{(final)} }| \kern-\nulldelimiterspace}{\Psi_f}\rangle
\strut_{ab} }{ \operatorname{Tr} \rho_{ab}^{(final)}},
\end{equation}
where $\rho_{ab}^{(final)}$ is the actual final state of the main
field modes $\hat a$ and $\hat b$, defined by Eqs.~(\ref{eqn79}),
(\ref{eqn74}).

Using Eqs.~(\ref{eqn74}), (\ref{eqn83}) and carrying out quite
straightforward mathematical calculations, one can obtain the
following expression for the fidelity of final state generation:
\begin{equation}
\label{eqn98} \begin{aligned} & F \approx 1- \sum_{ \{\Delta M_i \}
} \operatorname{Tr} \left\{ P_\bot \Delta\hat M_i
\rho_{ab}^{(final)} \right\} - {} \\ & {} - \frac{ \zeta}{\lambda
|\gamma|^2} |c_K|^2 \sum_{n_1} \operatorname{Tr} \left\{ P_\bot
|\Psi(n_1) \mathrel{\rangle \strut_{ab} \strut_{ab} \langle}
\Psi(n_1) | \right\},
\end{aligned}
\end{equation}
where $ P_\bot = 1- \ |{\Psi_f}\mathrel{\rangle{ \strut_{ab}
\strut_{ab} \vphantom{\Psi_f \Psi_f}}\langle
\kern-\nulldelimiterspace}{\Psi_f}|$.

For further consideration it is convenient to introduce unnormalized
states $|\Psi_f^{(s)} \rangle_{ab} = \sum_{n=0}^K n^s c_n |\alpha
e^{i \chi n} \rangle_a |\beta e^{i \chi n} \rangle_b$, which are
useful, for example, when exponent is decomposed in
Eq.~(\ref{eqn85}). Then it is quite easy to show that
\begin{equation}
\label{eqn99} \operatorname{Tr} \left\{ P_\bot \Delta \hat M_0 (\hat
a, \hat b) \rho_{ab}^{(final)} \right\} = 2 \eta_2 \left|
\strut_{ab} \langle{\Psi_f}\mathrel{|{ \vphantom{\Psi_f
\Psi_f^{(1)}}}.
\kern-\nulldelimiterspace}{\Psi_f^{(1)}}\rangle\strut_{ab}
\right|^2,
\end{equation}
where we suppose that the term proportional to $\eta_1$ vanishes due
to correct phase compensation.

Representing states of the form $|\alpha e^{i \chi_{ac} n}
\rangle_a$ and $|\beta e^{i \chi_{bc} n} \rangle_b$ as
\begin{equation}
\label{eqn88} |\alpha e^{i \chi_{ac} n} \rangle_a = e^{i \Delta
\chi_{ac} n \hat a^+ \hat a } |\alpha e^{i \chi n} \rangle_a
\end{equation}
and decomposing exponents in power series in this expression, one
can simplify the term in the sum in Eq.~(\ref{eqn98}), corresponding
to deviations of nonlinearity strengthes from their expected value
$\chi$:
\begin{equation}
\label{eqn100}\begin{aligned} \operatorname{Tr}& \left\{ P_\bot
\Delta \hat M_\chi (\hat a, \hat b)  \rho_{ab}^{(final)} \right\} =
{}\\{} = {} & \langle{\Psi_f^{(1)}}\mathrel{| \left( \Delta
\chi_{ac} \hat a^+ \hat a + \Delta \chi_{bc} \hat b^+ \hat b
\right)^2 | \kern-\nulldelimiterspace}{\Psi_f^{(1)}}\rangle - {} \\
& {} - \left| \langle{\Psi_f^{(1)}}\mathrel{| \left( \Delta
\chi_{ac} \hat a^+ \hat a + \Delta \chi_{bc} \hat b^+ \hat b \right)
| \kern-\nulldelimiterspace}{\Psi_f}\rangle \right|^2.
\end{aligned}
\end{equation}

Analytical expression for the remaining term in the sum can be found
in two limiting cases:
\begin{equation}
\label{eqn90} \operatorname{Tr} \left\{ P_\bot \hat M_2^{(1)} (\hat
a) \rho_{ab}^{(final)} \right\} \approx |\alpha|^2 \chi^2 \left(
\Lambda |\gamma|^2 + \Lambda^2 |\gamma|^4 \right)
\end{equation}
for $|\alpha|^2 \chi^2 \ll 1$ and
\begin{equation}
\label{eqn16}\operatorname{Tr} \left\{ P_\bot \hat M_2^{(1)} (\hat
a) \rho_{ab}^{(final)} \right\} \approx  \Lambda |\gamma|^2
\end{equation}
for $|\alpha|^2 \chi^2 \gg 1$.

Eqs.~(\ref{eqn98}), (\ref{eqn99}), (\ref{eqn100})--(\ref{eqn16})
provide the expression for the final state generation fidelity for
any desired final state $|\Psi_f \rangle_{ab}$ described by
Eq.~(\ref{eqn3}).

\subsection{Estimation of the system parameters, required for
protocol implementation}

Further simplification of the derived above equations for the final
state fidelity can be carried out in special cases, considered in
Section~\ref{sec:special}. We discuss the states with maximal
entanglement, possible for a scheme with fixed number of detectors
($K=1$ and $K=2$). On the basis of explicit expressions for the
fidelity of such states generation we find conditions, that
must be imposed on the system parameters in order to obtain
sufficiently high fidelity of final state generation.

For most of physical systems, suitable for implementation of the
protocol, assumptions that $\chi \ll 1$ and $|\alpha|^2 \chi^2 \ll
1$ are valid. In this paper we discuss analytical results
obtained under these assumptions only (however, the opposite
limiting case $|\alpha|^2 \chi^2 \gg 1$ can also be described
analytically).

For maximally entangled state, generated by the scheme with $K=1$
detector (Eqs.~(\ref{eqn11}), (\ref{eqn12})), we obtain
\begin{equation}
\label{eqn101} \operatorname{Tr} \left\{ P_\bot \Delta \hat M_0
(\hat a, \hat b) \rho_{ab}^{(final)} \right\} = \frac{ \eta_2}
{|\alpha|^2 \chi^2},
\end{equation}
\begin{equation}
\label{eqn102} \begin{aligned} \operatorname{Tr}& \left\{ P_\bot
\Delta \hat M_\chi (\hat a, \hat b)  \rho_{ab}^{(final)} \right\} =
{}\\{} = {} & \frac{2 |\alpha |^2 \left(\varepsilon_{{ac}}+
\varepsilon_{{bc}}\right)^2+\left(\varepsilon _{{ac}}-
\varepsilon_{{bc}}\right)^2}{4 }
\end{aligned}
\end{equation}
and
\begin{equation}
\label{eqn103} \begin{aligned}  \frac{ \zeta}{\lambda |\gamma|^2}
|c_K|^2 \sum_{n_1} \operatorname{Tr} \left\{ P_\bot |\Psi(n_1)
\mathrel{\rangle \strut_{ab} \strut_{ab} \langle} \Psi(n_1) |
\right\} = {} \\ {} =  \frac{\zeta }{2 \lambda |\gamma|^2 |\alpha|^2
\chi^2 },
\end{aligned}
\end{equation}
where $\varepsilon_{ac,bc} = \Delta \chi_{ac,bc} / \chi$ are
relative inaccuracies of the nonlinear interaction strengthes; we
assumed for simplicity that $|\alpha| = |\beta|$ and $q=1/\sqrt{K}$.

For the scheme with $K=2$ detectors and coefficients $c_n$ described
by Eq.~(\ref{eqn13}) the form of Eqs.~(\ref{eqn101})--(\ref{eqn103})
remains the same, but expressions in the right hand side of
Eqs.~(\ref{eqn101}), (\ref{eqn102}) get numerical factor 2.

The finally obtained expression for decrease of the final state
fidelity contains 6 distinct terms, corresponding to different types
of processes in the system: Eq.~(\ref{eqn101}) corresponds to 3
terms, proportional to $\Delta \phi^2$, $\Lambda_1$ and $\Lambda_2$
(see Eq.~(\ref{eqn87}) for $\eta_2$); Eq.~(\ref{eqn102}) and
Eq.~(\ref{eqn103}) provide expressions for the terms, describing
nonlinearity strength inaccuracies and nonideality of photodetectors
respectively; Eq.~(\ref{eqn90}) corresponds to decoherence of the
mode $\hat c$ that can be described by effective discrete phase
errors in the mode $\hat a$. In order to estimate parameters values,
suitable for final state generation with sufficient fidelity, we
require each of the discussed 6 terms to be not greater than some
small value $\epsilon$, $\epsilon \ll 1$ (then the fidelity will be
not less than $F\ge 1- 6 \epsilon$). The obtained 6 inequalities can
be used for finding 6 independent system parameters.

For this purpose we divide parameters, describing the system, into
four groups:
\begin{enumerate}
\item parameters $\zeta$, $\lambda$, $|\alpha|$ characterize exploited
"local" equipment (photodetectors, maximal field intensities
providing small decoherence during local operations) and are
supposed to be fixed by characteristics of existing equipment; in
numerical estimations we assume that dark count probability has the
values $\zeta = 10^{-8}$ when detectors efficiency equals $\lambda =
10^{-2}$ and $\zeta = 10^{-6}$ for $\lambda = 10^{-1}$ (such values
can be achieved for InGaAs/InP\- photodetectors
\cite{stucki-2008,InGaAs-detectors-2010}); we also assume that
$|\alpha|^2 \sim 10$;
\item parameters $\chi$, $\Delta \chi_{ac}$, $\Delta \chi_{bc}$,
$\Lambda_1$ characterize local cross-Kerr interaction; we use
the inequalities to estimate these parameters values and to find out
whether such nonlinear interaction can be realized experimentally;
\item parameters $\Lambda$, $\Delta \phi^2$, $\Lambda_2$
characterize properties of the quantum channel and local resonator
and determine maximal distance of entanglement generation;
\item the ancillary field amplitude $|\gamma|$ can be changed and is
chosen so as to provide maximal success probability for sufficiently
high fidelity of the final state generation.
\end{enumerate}

Therefore, the discussed inequalities, providing sufficient
fidelity, can be expressed in the following way (for $K=1$; for
$K=2$ the conditions are the same except for numerical factor 1/2 in
conditions 2, 3, 4, 6):
\begin{equation}
\label{eqn21} \left\{
\begin{aligned}
&\Lambda < \displaystyle\frac{2 \epsilon^2 \lambda}{\zeta}, \\
&\Lambda_2 < 2 \epsilon, \\
&\Delta \phi^2< |\alpha|^2 \chi^2  \epsilon, \\
&\Lambda_1 < \displaystyle \frac{3}{2}\, \epsilon, \\
&|\gamma|^2 \lesssim  \displaystyle \frac{\epsilon}{|\alpha|^2
\chi^2 \Lambda }, \\
&\varepsilon_{ac}^2, \varepsilon_{bc}^2 < \displaystyle
\frac{\epsilon}{2 |\alpha|^2 }.
\end{aligned} \right.
\end{equation}

The first of the conditions limits acceptable losses in the quantum
channel. This limitation is fundamental for optical methods of
information processing when dark counts of photodetectors are
present (see e.g. comments in Ref.~\cite{lutkenhaus-2006}). For
instance, for the considered above parameters and the desired
fidelity $F=0.9$ the maximal acceptable attenuation of the channel
is limited by values $(20\div 28)\mbox{ dB}$, which for optical
fiber with attenuation 0.20~dB/km correspond to maximal distances
about $L_{max} \approx (100 \div 140) \mbox{ km}$. However, it
should be noted, that entanglement generation at such distances
requires quite long storage of the field $\hat a$ in Alice's local
resonator. The value of the resonator finesse, required for
preserving sufficient state fidelity and defined on the basis of the
second condition of Eq.~(\ref{eqn21}), is about $10^{12}$ in this
case. Even greater values have already been predicted theoretically
for crystalline whispering gallery mode resonators
\cite{high-Q_cavity-74-063806}. Experimentally demonstrated
high-quality resonators are characterized by values up to $10^{9}
\div 10^{11}$
\cite{high-Q_cavity-74-063806,high-Q_cavity-71-013817}. Therefore,
we believe that the protocol will be more suitable for efficient
entanglement generation when the best available quantum channel
connecting Alice's and Bob's sites is lossier than optical fiber.

The third condition provides lower bound on the nonlinearity value.
For $\Delta \phi \sim 10^{-3} \div 10^{-2}$ and the considered above
parameters the minimal nonlinearity strength is $\chi_{min} \sim
10^{-3} \div 10^{-2}$. Such values has already been predicted in
existing systems for the case of precise radiation focusing
\cite{imamoglu1, imamoglu2,
sinclair,kang_zhu,kang_zhu-1,li_yang-2008}. It should be noted, that
not only nonlinearity strength, but also acceptable signal
attenuation during cross-Kerr interaction is limited (the fourth
condition of Eq.~(\ref{eqn21}) leads to requirement $\Lambda_1 <
0.025$).

\begin{figure}[t]
\begin{center}
\begin{tabular}{cc}
\textbf{(a)} & \\ &  \includegraphics[scale=0.8]{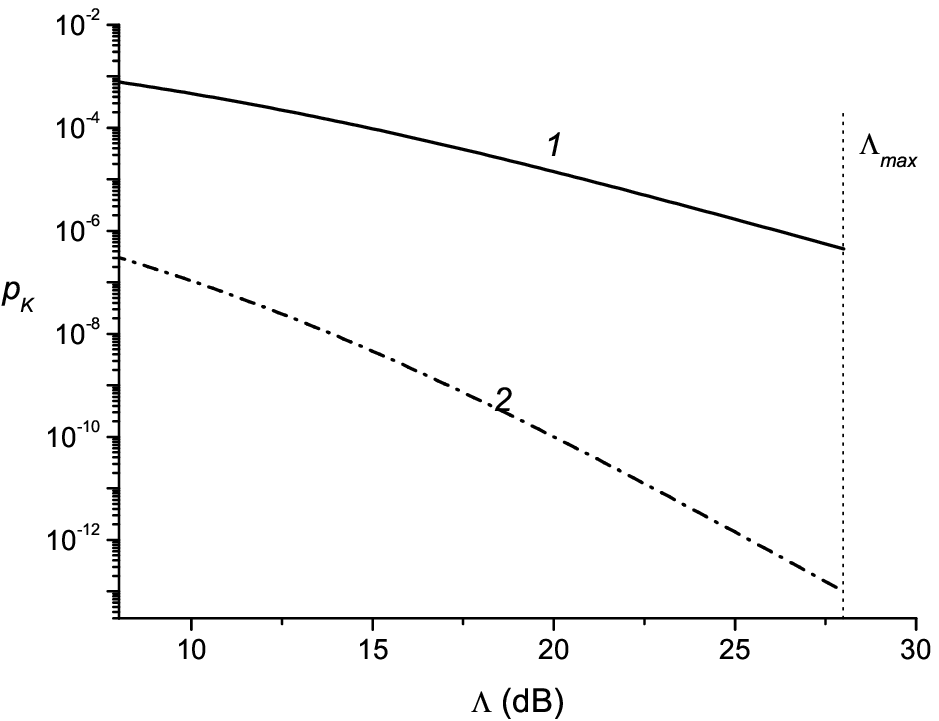}  \\ \\
\textbf{(b)} & \\ & \includegraphics[scale=0.8]{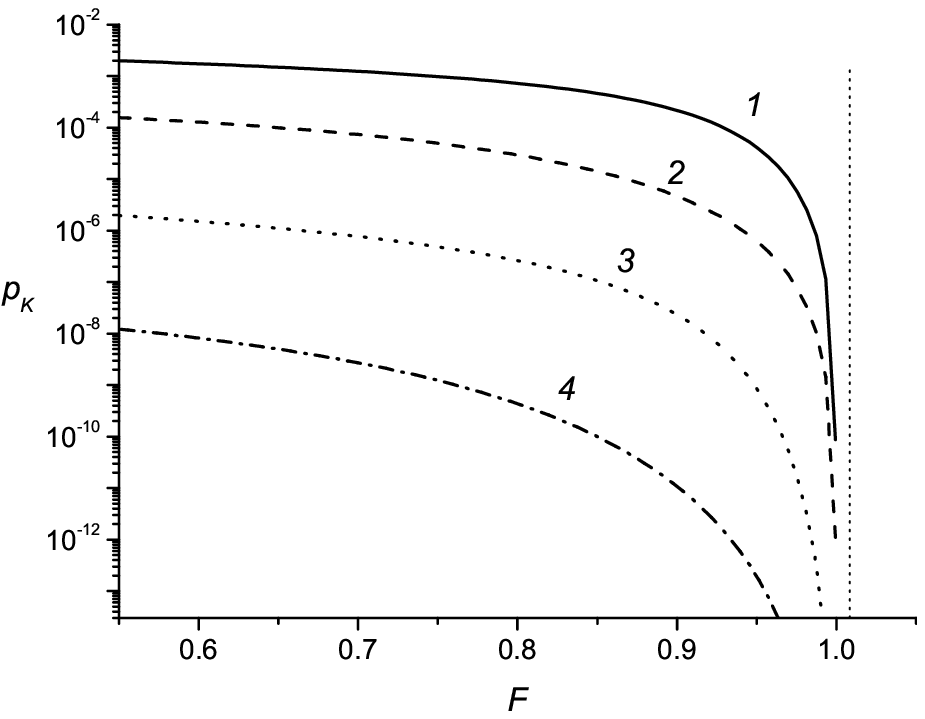} \\ \\
\end{tabular}
\caption{\textbf{(a)} Dependence of one-run success probability
$p_K$ on the relative losses rate $\Lambda$ for the protocol with
$K$ photodetectors for fixed fidelity of final state generation
$F=0.9$: $K=1$ (1), $K=2$ (2). Dotted line corresponds to the limit
of entanglement generation because of dark counts of detectors.
\textbf{(b)} Dependence of one-run success probability $p_K$ on the
desired fidelity $F$ for the protocol with $K$ photodetectors and
different losses rates $\Lambda$: $K=1$, $\Lambda=14\mbox{ dB}$ (1),
$K=1$, $\Lambda=28\mbox{ dB}$ (2), $K=2$, $\Lambda=14\mbox{ dB}$
(3), $K=2$, $\Lambda=28\mbox{ dB}$ (4).} \label{fig4}
\end{center}
\end{figure}

The fifth condition of Eq.~(\ref{eqn21}) limits maximal probe beam
intensities and determines the maximal possible one-run success
probability for the protocol, described by Eq.~(\ref{eqn82}). For
the protocol with $K=1$ detector the success probability is
sufficiently large for all losses values not exceeding the limit
determined by the first condition of Eq.~(\ref{eqn21})
(Fig.~\ref{fig4}). For $K = 2$ generation of the desired final state
can be implemented without too large number of ancillary field
transmissions for losses not more than approximately 14~dB (which
correspond to the distances up to 70~km in optical fiber).

The maximal acceptable relative inaccuracies of nonlinearity
strengthes $\varepsilon_{ac}$, $\varepsilon_{bc}$, defined by the
last condition of Eq.~(\ref{eqn21}), are equal to 0.09 for $K=1$ and
to 0.06 for $K=2$.

Such parameters values can be achieved in real systems, and,
therefore, our calculations prove applicability of the protocol for
entanglement generation between sites, separated by lossy media,
using contemporary experimental equipment.

\section{Conclusions}

To summarize, in the present work we have proposed a protocol for
creating a wide class of qudit-type states (including entangled
states) with arbitrary dimensionality in continuous variable system
using weak cross-Kerr nonlinearity, linear beamsplitters, detectors
not resolving photon numbers, and sources of coherent states.

The method of entanglement generation is based on using an ancillary
field mode, transmitted from Alice's site to Bob's one through lossy
quantum channel. Weak nonlinear interaction of the mode with the
main field modes possessed by Alice and Bob leads to creation of a
weakly entangled 3-modes state. The main problem, solved in our work
is designing a scheme for the ancillary mode state measuring leading
to probabilistic entanglement enhancement and transforming the "raw"
weakly correlated state into highly entangled final one. The found
POVM measurement is shown to be implementable with linear optics and
photodetectors, not resolving photon numbers, on the basis of
elimination measurements. The equation, defining parameters of the
detection scheme for a given desired final state in a unique way, is
also derived in our paper.

Our calculations prove applicability of the proposed protocol in the
case of realistic photodetectors with limited efficiency and dark
counts, nonlinear Kerr interaction with decaying modes, uncertainty
of coupling constants, and lossy quantum channel. It has been shown
that the protocol can be used for creating quantum states with
entanglement higher than unity and, therefore, in certain cases
corresponds to more effective use of resources than can be achieved
for protocols based on entangling qubit systems. The fidelity of
final state generation $F=0.9$ can be achieved when a quantum
channel with losses rate up to $(20\div 28)\mbox{ dB}$ is available
(it corresponds to distances up to 140~km for optical fiber).
Required cross-Kerr nonlinearity is $\chi \ge \chi_{\min}\sim
10^{-3} \div 10^{-2}$ and can be created using contemporary
equipment.

\subsection{Acknowledgements}

The work has been supported partially by the European Commission via
project "Engineering Quantum Information in Nanostructured Diamond".

\appendix

\section{Elimination of coherent states as POVM measurements}
\label{app1}

Here we provide mathematical description of system state
transformation when measurements, eliminating coherent states
$\{|\gamma_j \rangle_c \}$, are implemented. The considered
measurement scheme is shown in Fig.~\ref{fig2}(d). The mode $\hat c$
is mixed at linear beamsplitters with $K$ reference modes $\hat
d_j$, prepared in coherent states $|\tilde \gamma_j \rangle_{d_j}$
by splitting the initial coherent state of the mode $\hat d$. Then
photodetectors $D_j$ determine presence of photons in the modes
$\hat d_j$.

Let the state of modes $\hat a$, $\hat b$ and $\hat c$ before
elimination measurements be described by density matrix
$\rho_{abc}^{(in)}$. Taking into account that coherent states
represent a natural basis for describing linear optical
transformations, it is convenient to decompose the input density
matrix in terms of coherent states of the mode $\hat c$
\begin{equation}
\label{eqn39} \rho_{abc}^{(in)} = \int \hat P_{ab}(\gamma_x)
|{\gamma_x}\mathrel{\rangle{ \strut_c \strut_c \vphantom{\gamma_x
}}\langle \kern-\nulldelimiterspace}{\gamma_x}| d^2\gamma_x,
\end{equation}
where $\hat P_{ab}(\gamma_x)$ is an operator-valued function of
variable $\gamma_x$ (acting as an operator on the modes $\hat a$ and
$\hat b$), analogous to Glauber function of one-mode field.

The state of the expanded system, composed by the main modes $\hat
a$ and $\hat b$, the ancillary mode $\hat c$ and reference modes
$\hat d_j$, after mixing the ancillary mode $\hat c$ with reference
modes $\hat d_j$ at beamsplitters is described by the following
density matrix:
\begin{widetext}
\begin{equation}
\label{eqn38} \rho_{abcd_1...d_K} = \hat T_{c d_K} ... \hat T_{c
d_1} \left( \rho_{abc}^{(in)} \otimes \left|{\tilde
\gamma_1}\mathrel{\left\rangle {\vphantom{\tilde \gamma_1}_{d_1
d_1}} \right\langle \kern-\nulldelimiterspace}{\tilde
\gamma_1}\right| \otimes ... \otimes  \left|{\tilde
\gamma_K}\mathrel{\left\rangle {\vphantom{\tilde \gamma_K}_{d_K
d_K}} \right\langle \kern-\nulldelimiterspace}{\tilde
\gamma_K}\right| \right) \hat T_{c d_1}^+ ... \hat T_{c d_K}^+,
\end{equation}
\end{widetext}
where $\hat T_{cd_j}=\exp\left\{i\theta_j \left(\hat c^+ \hat d_j+
\hat c \hat d_j^+ \right) \right\}$ are the unitary operators of
field transformation by beamsplitters (the quantity $\theta_j$ is
related to transmittance as $T_j  = \cos^2 \theta_j$).

Using Eq.~(\ref{eqn39}) one can represent the density matrix
(\ref{eqn38}) in the form
\begin{equation}
\label{eqn40} \begin{aligned} & \rho_{abcd_1...d_K} = {} \\ &{} =
\int \hat P_{ab}(\gamma_x) |{\Psi(\gamma_x)}\mathrel{\rangle{
\strut_{cd_1...d_K} \strut_{cd_1...d_K} \vphantom{\Psi(\gamma_x)
\Psi(\gamma_x)}}\langle \kern-\nulldelimiterspace}{\Psi(\gamma_x)}|
d^2\gamma_x,
\end{aligned}
\end{equation}
where $|\Psi(\gamma_x)\rangle_{cd_1...d_K} = \hat T_{c d_K} ... \hat
T_{c d_1} |\gamma_x \rangle_c |\tilde \gamma_1 \rangle_{d_1} ...
|\tilde \gamma_1 \rangle_{d_K}$.

The values of transmittances $T_j$ of the beamsplitters $BS_j$
(Eq.~(\ref{eqn35})) and amplitudes $\tilde \gamma_j$ of the
reference modes $\hat d_j$ (Eq.~(\ref{eqn75})) are chosen in such a
way, that amplitude of coherent state $|\gamma_x \rangle_c$ of the
ancillary mode is split in equal parts between the modes $\hat d_j$,
and each of the reference modes effectively undergoes coherent
displacement $- \gamma_j$:
\begin{equation}
\label{eqn41}\begin{aligned} &|\Psi(\gamma_x)\rangle_{cd_1...d_K} =
| \gamma_0' (\gamma_x) \rangle_c \otimes \\&\otimes \left|i q \cdot
\left(\gamma_x - \gamma _{1}\right) \right\rangle_{d_1} ... \left| i
q \cdot \left(\gamma_x - \gamma _{K}\right)
\right\rangle_{d_K},\end{aligned}
\end{equation}
where $\gamma_0' (\gamma_x)$ is a linear function of the amplitude
$\gamma_x$ of the ancillary mode coherent state, equal to
\begin{equation}
\label{eqn44} \gamma_0 ' (\gamma_x) = \sqrt{ 1-K q^2} \; \gamma_x +
\sqrt{\frac{1-\delta}{\delta}}\; q \sum_{j=1}^K \gamma_K.
\end{equation}

Measuring presence of photons in the modes $\hat d_j$ by the
photodetectors can be described by a set of pairs of projective
operators
\begin{equation}
\label{eqn29} \hat P_{j-} =
 \left|{0}\mathrel{\left\rangle
{\vphantom{0}_{d_j d_j}} \right\langle
\kern-\nulldelimiterspace}{0}\right|\quad \mbox{and} \quad \hat
P_{j+} = 1- \hat P_{j-},
\end{equation}
describing absence and presence of photons in corresponding field
mode respectively.

The final state of modes $\hat a$, $\hat b$ and $\hat c$ after
carrying out the measurements and discarding reference modes $\hat
d_j$ is described by density matrix
\begin{widetext}
\begin{equation}
\label{eqn31} \rho_{abc}^{(out)} =  \int \hat P_{ab}(\gamma_x)
|{\gamma_0'(\gamma_x)}\mathrel{ \rangle{ \strut_c \strut_c
\vphantom{\gamma_0' (\gamma_x)}}\langle
\kern-\nulldelimiterspace}{\gamma_0'(\gamma_x)}|\cdot \prod_{j=1}^K
\operatorname{Tr}_{d_j} \left\{ \hat P_{j \pm} |{i q \cdot
\left(\gamma_x - \gamma _{j}\right)}\mathrel{\rangle{ \strut_{d_j}
\strut_{d_j} \vphantom{i q \cdot \left(\gamma_x - \gamma
_{K}\right)}}\langle \kern-\nulldelimiterspace}{i q \cdot
\left(\gamma_x - \gamma _{j}\right)}|  \right\} \cdot d^2\gamma_x,
\end{equation}
where the type of used projector ("+" or "-") depends on the
obtained measurement outcome. This expression can be simplified
using the following relation:
\begin{equation} \label{eqn43} \begin{aligned}
\operatorname{Tr}_{d_j} \left\{ \hat P_{j \pm} |{i q \cdot
\left(\gamma_x - \gamma _{j}\right)}\mathrel{\rangle{  \strut_{d_j}
\strut_{d_j} \vphantom{i q \cdot \left(\gamma_x - \gamma
_{K}\right)}}\langle \kern-\nulldelimiterspace}{i q \cdot
\left(\gamma_x - \gamma _{j}\right)}|  \right\} = \sum_{n_j\in
\Omega_\pm} \left| \strut_{d_j}
\left\langle{n_j}\mathrel{\left|{\vphantom{n i q \cdot
\left(\gamma_x - \gamma _{j}\right)}}\right.
\kern-\nulldelimiterspace}{i q \cdot \left(\gamma_x - \gamma
_{j}\right)}\right\rangle\strut_{d_j}  \right|^2 = {} \\ {} =
\sum_{n_j\in \Omega_\pm} \frac{q^{2 n_j}}{n_j!} \left| \gamma_x -
\gamma_j \right|^{2 n_j} e^{- q^2\left| \gamma_x - \gamma_j
\right|^2},
\end{aligned}
\end{equation}
where $\Omega_{-} = \{0\}$ and $\Omega_+ = \{n\mathrel{|}n\ge 1\}$
are the sets of photon numbers, corresponding to the measurement
outcomes "-" (absence of photons detected) and "+" (photocount
obtained) respectively.

Introducing superoperator, which describes measurement-invariant
part of the ancillary mode state transformation by the definition
\begin{equation}
\label{eqn32} \hat M \colon |{\gamma_x}\mathrel{\rangle{ \strut_c
\strut_c \vphantom{\gamma_x }}\langle
\kern-\nulldelimiterspace}{\gamma_x}| \mapsto
|{\gamma_0'(\gamma_x)}\mathrel{\rangle{ \strut_c \strut_c \vphantom
{\gamma_0'(\gamma_x)}} \langle
\kern-\nulldelimiterspace}{\gamma_0'(\gamma_x)}| \exp \biggl( -q^2
\sum_{j=1}^K \left| \gamma_x - \gamma_j \right|^2 \biggr),
\end{equation}
one can transform Eq.~(\ref{eqn31}) for the density matrix
$\rho_{abc}^{(out)}$ to the form
\begin{equation}
\label{eqn42} \rho_{abc}^{(out)} =  \hat M \left\{ \int \hat
P_{ab}(\gamma_x) |\gamma_x\mathrel{ \rangle{ \strut_c \strut_c
\vphantom{\gamma_x}\langle }
\kern-\nulldelimiterspace}{\gamma_x}|\cdot \prod_{j=1}^K
\sum_{n_j\in \Omega_\pm} \frac{q^{2 n_j}}{n_j!} \left| \gamma_x -
\gamma_j \right|^{2 n_j} \cdot d^2\gamma_x \right\} ,
\end{equation}

Then one can define operators, which act as follows
\begin{equation}
\label{eqn45} \hat B_{jn_j} = \frac{q^{n_j}}{\sqrt{n_j!}} \left(
\hat c - \gamma_j \right)^{n_j} \quad \colon \quad  |\gamma_x
\rangle_c \mapsto \frac{q^{n_j}}{\sqrt{n_j!}} \left( \gamma_x -
\gamma_j \right)^{n_j} |\gamma_x \rangle_c ,
\end{equation}
and transform the expression (\ref{eqn42}) for the final state
density matrix in the following way:
\begin{equation}
\label{eqn33}\begin{aligned} \rho_{abc}^{(out)} = \hat M \left\{
\int \hat P_{ab}(\gamma_x) \cdot \sum_{n_j \in \Omega_\pm} \hat
B_{Kn_K} ... \hat B_{1n_1} |\gamma_x\mathrel{ \rangle{ \strut_c
\strut_c \vphantom{\gamma_x}\langle }
\kern-\nulldelimiterspace}{\gamma_x}| \hat B_{1n_1}^+ ... \hat
B_{Kn_K}^+  \cdot d^2\gamma_x \right\} = {} \\ {} = \hat M \left\{
\sum_{n_j \in \Omega_\pm} \hat B_{Kn_K} ... \hat B_{1n_1}
\rho_{abc}^{(in)} \hat B_{1n_1}^+ ... \hat B_{Kn_K}^+ \right\}.
\end{aligned}
\end{equation}
\end{widetext}

Eq.~(\ref{eqn33}) describes transformation of the system density
matrix by elimination measurements in the form, similar to the one,
corresponding to POVM measurements. It should be noted, however,
that, the standard normalization condition is satisfied only for
complete state transformation by the measuring setup (including
action of superoperator $\hat M$ and $K$ operators $\hat
B_{1n_1}$,~..., $\hat B_{Kn_K}$) rather than for single state
elimination.

Operators $\hat B_{jn_j}$ for $n_j>0$ correspond to the definition
Eq.~(\ref{eqn7}) of operators, describing successful elimination of
the state $|\gamma_j \rangle_c$:
\begin{equation}
\label{eqn84} \hat B_{jn_j} |\gamma_j \rangle_c =
\frac{q^{n_j}}{\sqrt{n_j!}} \left( \gamma_x - \gamma_j \right)^{n_j}
|\gamma_x \rangle_c = 0,\; n_j=1,2,...
\end{equation}
Therefor, returning to the notations of Eqs.~(\ref{eqn7}),
(\ref{eqn23}), we can define operators $\hat A_{0n_j}
(|\gamma_j\rangle_c)$, eliminating coherent state
$|\gamma_j\rangle_c$, as
\begin{equation}
\label{eqn46} \hat A^{(n_j)} _{|\gamma_j\rangle_c} = \hat B_{jn_j}
\equiv \frac{q^{n_j}}{\sqrt{n_j!}} \left( \hat c - \gamma_j
\right)^{n_j}, \quad n_j=1,2,...
\end{equation}
These operators satisfy the conditions, provided by
Eqs.~(\ref{eqn7}), (\ref{eqn23}):
\begin{equation}
\label{eqn47} \hat A^{(n_j)} _{|\gamma_j\rangle_c}
|\gamma_j\rangle=0,
\end{equation}
\begin{equation}
\label{eqn48} \left[\hat A^{(n_i)} _{|\gamma_i\rangle_c}, \hat
A^{(n_j)} _{|\gamma_j\rangle_c}  \right]=0 \quad \mbox{for all }
n_i,\,n_j.
\end{equation}

According to Eq.~(\ref{eqn33}), the final state of the system in the
case of successful outcome of elimination of all the states $\{
|\gamma_j \rangle_c \}$ is described by density matrix
\begin{widetext}
\begin{equation} \label{eqn49a} \rho_{abc}^{(out)} = \hat M \left\{
\sum_{n_j\ge 1} \hat A^{(n_K)} _{|\gamma_K\rangle_c} ...  \hat
A^{(n_1)} _{|\gamma_1\rangle_c} \rho_{abc}^{(in)} \left( \hat
A^{(n_1)} _{|\gamma_1\rangle_c} \right)^+ ... \left( \hat A^{(n_K)}
_{|\gamma_K\rangle_c} \right)^+ \right\}.
\end{equation}

If the input state is $|\Psi_1 \rangle_{abc} = \hat D_c \left( \hat
F_{ab} \gamma \right) |\alpha \rangle_a |\beta \rangle_b |0
\rangle_c$ (see Eqs.~(\ref{eqn1}), (\ref{eqn55})), the expression
for the final state can be rewritten as
\begin{equation}
\label{eqn57} \rho_{abc}^{(out)} = \hat M \left( \sum_{n_j\ge 1}
\frac{(q|\gamma|)^{2 n_1 + ... + 2 n_K}}{n_1!...n_K!}  \left( \hat
F_{ab} - \frac{\gamma_K}{\gamma}\right)^{n_K} ... \left( \hat F_{ab}
- \frac{\gamma_1}{\gamma}\right)^{n_1}
|{\Psi_1}\mathrel{\rangle{\strut_{abc} \strut_{abc}
\vphantom{\Psi_1}}\langle \kern-\nulldelimiterspace}{\Psi_1}| \left(
\hat F_{ab}^+ - \frac{\gamma_1^\ast}{\gamma^\ast}\right)^{n_1} ...
\left( \hat F_{ab}^+ -
\frac{\gamma_K^\ast}{\gamma^\ast}\right)^{n_K} \right),
\end{equation}
\end{widetext}
where Eqs.~(\ref{eqn46}), (\ref{eqn58}) were taken into account. For
small ancillary field amplitudes $|\gamma| \ll 1$ the main
contribution to the final state density matrix is made by the term
with $n_1=...=n_K=1$ (the most probable case of successful
elimination of the states $\{|\gamma_j\rangle\}$ corresponds to
detection of exactly 1 photon by each of the detectors). After
discarding mode $\hat c$, the final state of modes $\hat a$ and
$\hat b$ is described by density matrix
\begin{widetext}
\begin{equation}
\label{eqn59} \rho_{ab} = \operatorname{Tr}_c \rho_{abc}^{(out)} =
q^{2K} |\gamma|^{2K} |{\Psi_f'}\mathrel{\rangle{ \strut_{ab}
\strut_{ab} \vphantom{\Psi_f' }}\langle
\kern-\nulldelimiterspace}{\Psi_f'}| + O \left( |\gamma|^{2K+2}
\right),
\end{equation}
where
\begin{equation}
\label{eqn60} |\Psi_f' \rangle_{ab}  = \left( \hat F_{ab} -
\frac{\gamma_K}{\gamma}\right) ... \left( \hat F_{ab} -
\frac{\gamma_1}{\gamma}\right) |\alpha \rangle_a |\beta \rangle_b =
\frac{1}{c_K} \sum_{n=0}^K c_n \hat F_{ab}^n |\alpha \rangle_a
|\beta \rangle_b = \frac{1}{c_K} |\Psi_f \rangle_{ab};
\end{equation}
\end{widetext}
we have taken into account that amplitudes $\{\gamma_j\}$ are roots
of Eq.~(\ref{eqn8}); $|\Psi_f \rangle_{ab}$ is the desired final
state, described by Eq.~(\ref{eqn3}).

\end{document}